\documentclass[aps,prb,onecolumn,superscriptaddress]{revtex4-1}

\usepackage{graphicx}
\usepackage{bm}
\usepackage{subfigure}
\usepackage{lineno}
\usepackage{hyperref}
\usepackage{amsfonts}
\usepackage{amssymb}
\usepackage{courier}
\usepackage{array}

\newcolumntype{P}[1]{>{\centering\arraybackslash}p{#1}}
\newcolumntype{L}[1]{>{\arraybackslash}p{#1}}

\begin{document}

\title{Upgrade to the MAPS neutron time-of-flight chopper spectrometer}

\author{R. A. Ewings}
\email[Russell Ewings ]{russell.ewings@stfc.ac.uk}
\affiliation{ISIS Pulsed Neutron and Muon Source,
        STFC Rutherford Appleton Laboratory,
        Harwell Campus, Didcot OX11 0QX, United Kingdom}

\author{J. R. Stewart}
\affiliation{ISIS Pulsed Neutron and Muon Source,
        STFC Rutherford Appleton Laboratory,
        Harwell Campus, Didcot OX11 0QX, United Kingdom}

\author{T. G. Perring}
\affiliation{ISIS Pulsed Neutron and Muon Source,
        STFC Rutherford Appleton Laboratory,
        Harwell Campus, Didcot OX11 0QX, United Kingdom}
\affiliation{London Centre for Nanotechnology, 17-19 Gordon Street, London WC1H 0AH, United Kingdom}

\author{R. I. Bewley}
\affiliation{ISIS Pulsed Neutron and Muon Source,
        STFC Rutherford Appleton Laboratory,
        Harwell Campus, Didcot OX11 0QX, United Kingdom}

\author{M. D. Le}
\affiliation{ISIS Pulsed Neutron and Muon Source,
        STFC Rutherford Appleton Laboratory,
        Harwell Campus, Didcot OX11 0QX, United Kingdom}

\author{D. Raspino}
\affiliation{ISIS Pulsed Neutron and Muon Source,
        STFC Rutherford Appleton Laboratory,
        Harwell Campus, Didcot OX11 0QX, United Kingdom}

\author{D. E. Pooley}
\affiliation{ISIS Pulsed Neutron and Muon Source,
        STFC Rutherford Appleton Laboratory,
        Harwell Campus, Didcot OX11 0QX, United Kingdom}

\author{G. \v{S}koro}
\affiliation{ISIS Pulsed Neutron and Muon Source,
        STFC Rutherford Appleton Laboratory,
        Harwell Campus, Didcot OX11 0QX, United Kingdom}

\author{S. P. Waller}
\affiliation{ISIS Pulsed Neutron and Muon Source,
        STFC Rutherford Appleton Laboratory,
        Harwell Campus, Didcot OX11 0QX, United Kingdom}

\author{D. Zacek}
\affiliation{ISIS Pulsed Neutron and Muon Source,
        STFC Rutherford Appleton Laboratory,
        Harwell Campus, Didcot OX11 0QX, United Kingdom}

\author{C. A. Smith}
\affiliation{ISIS Pulsed Neutron and Muon Source,
        STFC Rutherford Appleton Laboratory,
        Harwell Campus, Didcot OX11 0QX, United Kingdom}

\author{R. C. Riehl-Shaw}
\affiliation{ISIS Pulsed Neutron and Muon Source,
        STFC Rutherford Appleton Laboratory,
        Harwell Campus, Didcot OX11 0QX, United Kingdom}

\date{\today}

\begin{abstract}
The MAPS direct geometry time-of-flight chopper spectrometer at the ISIS pulsed neutron and muon source has been in operation since 1999 and its novel use of a large array of position-sensitive neutron detectors paved the way for a later generations of chopper spectrometers around the world. Almost two decades of experience of user operations on MAPS, together with lessons learned from the operation of new generation instruments, led to a decision to perform three parallel upgrades to the instrument. These were to replace the primary beamline collimation with supermirror neutron guides, to install a disk chopper, and to modify the geometry of the poisoning in the water moderator viewed by MAPS. Together these upgrades were expected to increase the neutron flux substantially, to allow more flexible use of repetition rate multiplication and to reduce some sources of background. Here we report the details of these upgrades, and compare the performance of the instrument before and after their installation, as well as to Monte Carlo simulations. These illustrate that the instrument is performing in line with, and in some respects in excess of, expectations. It is anticipated that the improvement in performance will have a significant impact on the capabilities of the instrument. A few examples of scientific commissioning are presented to illustrate some of the possibilities.
\end{abstract}

\maketitle


\section{Introduction}\label{s:intro}

MAPS is a direct geometry time-of-flight (TOF) chopper spectrometer, located at the ISIS pulsed neutron and muon source at the Rutherford Appleton laboratory, UK \cite{MAPS-tech}. When it was first built it was a trail-blazing instrument, being the first spectrometer at a pulsed source that was designed from the outset to utilize a large array of position-sensitive detectors (PSDs) for the study of single crystal samples. It was originally envisaged that MAPS would be used predominantly for studies of high-energy excitations, using neutrons with incident energies in the epithermal range. Its use evolved over time, and more recently a significant proportion of the beam time has been devoted to single crystal and powder excitation experiments involving the use of thermal neutrons, measuring excitations with energies as low as a few meV addressing a wide variety of scientific problems \cite{Princep-YIG,Stock-CFO,Dalla-Piazza_NPhys15,Qureshi-LiFeAs,Stock-FeTe,Oh_PRL13,Schmidiger_PRL13,Lake-PRL-13,Johnstone-PRL-12,Lipscombe-PRL-11,Headings_PRL10,Doubble-PRL-10,Lake-NatPhys-10,Walters-NPhys-09,Diallo-PRL-09,Lipscombe-PRL-09,Xu_NPhys09,Vignolle_NPhys07,Hayden-Nature,Tranquada-Nature,Perring_manganite,Hayden_CrV}.
In addition a recent trend has seen about one quarter of the beam time become devoted to catalysis and molecular spectroscopy experiments \cite{Parker-SpecActa-18,Brown-PRB-17,OMalley-ChemCom-17,Albers-carbon-16,Cavallari-PhysChemChemPhys-16,Parker-JM-Rev-16,Marques-bone-16,Warringham-JChemPhys,Albers-Pearlmans-15,Parker-review}.

Following the success of MAPS a second generation of direct geometry chopper spectrometers has been built at neutron sources around the world, such as MERLIN and LET at ISIS \cite{Bewley20061029,Bewley-LET}; ARCS, SEQUOIA and CNCS at SNS \cite{Abernathy-2012,Granroth-SEQ,Ehlers-CNCS}; 4-SEASONS, HRC and AMATERAS at J-PARC \cite{4seasons,Itoh-HRC,Nakajima-AMATERAS}; IN5 at ILL \cite{Ollivier-IN5}; and NEAT at HZB \cite{RUSSINA2017}. As well as employing refinements of the PSD concept, with much larger detector arrays, all of these instruments make use of supermirror neutron guides in the primary spectrometer between the source and the sample in order to transfer much more of the source brilliance to the sample. In contrast MAPS followed a more traditional design for its primary spectrometer, using neutron-absorbing boron carbide (B$_{4}$C) collimation, as guide technology was less well-advanced at the time of the construction of MAPS than it is today. Given the successful operation of this second generation of spectrometers it was considered necessary to retro-fit a supermirror guide on to MAPS to allow it to continue to address topical questions in a wide variety of scientific fields. In particular there was a need in the ISIS spectrometer suite for a modern high-resolution thermal neutron instrument, to bridge the gap between the cold neutron high-resolution spectrometer LET and the thermal neutron medium-resolution spectrometer MERLIN.


\section{Pre-upgrade instrument description}\label{s:pre-upgrade}

We first briefly summarize the principles of operation of MAPS. At ISIS the neutron beam is produced by spallation, in which a high energy proton pulse strikes a tungsten target every 20\,ms (on ISIS target station 1) to produce a pulse of neutrons. These high energy neutrons are then moderated. The moderator viewed by MAPS is a poisoned ambient temperature water moderator \cite{Skoro-moderators}. As with other instruments in this class, on MAPS the polychromatic neutron beam is monochromated by the use of a mechanical chopper, in this case a Fermi chopper, i.e. a rotating collimator made from alternating neutron-transmitting slits and neutron-absorbing slats \cite{Fermi-orig}. The monochromatic beam is then incident on a sample and the scattering angle and time-of-flight of the scattered neutrons are recorded at position-sensitive detectors. The arrival time of the scattered neutrons at the detectors (i.e. TOF) together with knowledge of the sample orientation and the distance between different beamline components can be used to reconstruct the sample's momentum and energy dependent scattering function $S(\mathbf{Q},E)$.

\begin{figure*}[t]
\includegraphics*[scale=0.65,angle=0]{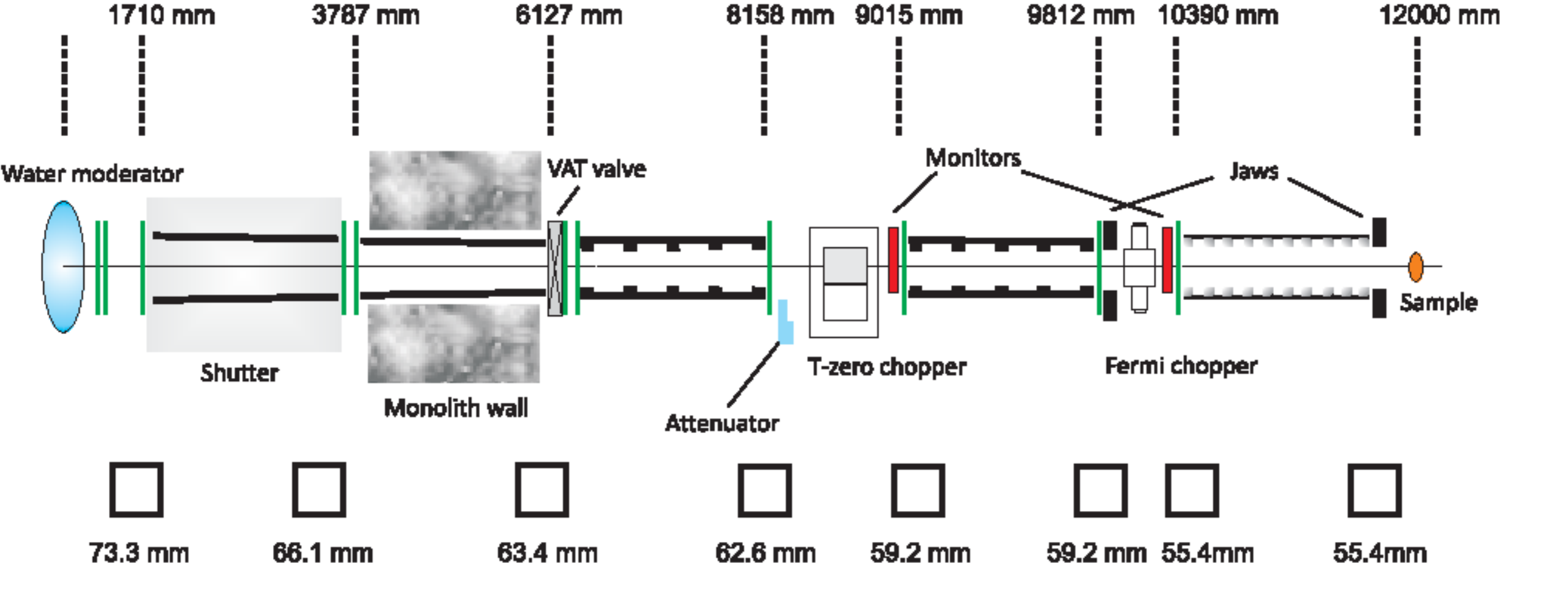}
\centering \caption{Schematic cross-section of the MAPS primary spectrometer before the upgrade (beam direction left to right). Distances of each component from the moderator are given in mm at the top of the figure, and beam apertures (square in all cases) at these components are given at the bottom of the figure. Green vertical lines indicate the position of thin aluminium windows.} \label{fig:layout_collimation}
\end{figure*}

A schematic of MAPS pre-upgrade is shown in fig. \ref{fig:layout_collimation}. Moving downstream from the target station (left to right on the schematic) the beamline first comprised a heavy shutter over two meters long which used to contain collimation made from sintered B$_{4}$C. In the wall of the target station monolith there was an insert section of collimation that consisted of a heavy steel housing filled with Borax (sodium tetraborate) and resin. The collimation tube inside this was lined with pressed B$_{4}$C powder and resin mixture formed into circular disks with a square cut-out in the middle that defined the beam aperture. At the downstream end of the insert there was a VAT vacuum valve \cite{VAT-ref} that was used to isolate the target station and monolith from the experimental hall. Next there was a further section of collimation following a similar design to that in the insert. This was followed by a large pit housing a chopper, known as the T-zero chopper, which was typically run at 50\,Hz, blocking the line of sight to the target station at the instant in time when the proton pulse strikes the target ($t =0$), in order to reduce the background measured in the detectors at later times. At the upstream end of the T-zero chopper pit there was a beam attenuator, consisting of a movable square piece of perspex that was slightly larger than the beam aperture. The first neutron beam monitor was placed at the downstream end of the T-zero chopper pit. The monitors were the scintillating glass type \cite{Monitor-1,Monitor-2}, widely used at ISIS \cite{Monitor-3,Monitor-4} \footnote{In this case a total of fifteen GS20\textsuperscript{\tiny\textregistered} cubic beads, 250\,$\mu m$ in size, are arranged in a net pattern to sample the neutron beam. The light from the GS-beads is collected using a photo-multiplier tube and the signal processed using electronics developed in-house to discriminate between neutron and gamma events based on the signal pulse height.}. There then followed another short section of collimation, and then a pit containing the Fermi chopper, at the downstream end of which was another monitor of the same design as the first. There were three Fermi chopper rotors available for use, with manual change-over between then taking approx. one hour. The rotor that gave the lowest energy resolution but highest flux was the so-called `sloppy chopper', which had slits with a radius of curvature of 1.3\,m, transmitting slit width 2.9\,mm and absorbing slat width 0.53\,mm. The rotor that gave the highest resolution but the lowest flux (typically) was the so-called `A chopper', which had slits with the same radius of curvature as the `sloppy chopper' and slats of the same width, but a slit width of 1.09\,mm. An intermediate resolution and flux rotor, the `B chopper' also had the same slat width, but had slits of width 1.81\,mm and a radius of curvature of the slits of 0.92\,m.

All of the collimation from the insert to the piece between the T-zero and Fermi choppers was evacuated using a vacuum circuit common to half of the instruments on the target station. The shutter was not under vacuum but rather it was continuously purged with helium gas. The final section of collimation shared a common vacuum with the sample and detector tank, and had a more sophisticated layout for the B$_{4}$C-resin mix pieces. As shown in the figure, they followed a sawtooth shape in order to prevent a line of sight from the sample to any of the walls of the collimation tube. The downstream end of each sawtooth was further faced with a piece of sintered B$_{4}$C. This was done to minimize the possibility that neutrons inelastically scattered by the resin in the B$_{4}$C-resin mix would arrive at the sample and then give rise to spurions (`spurion' is a colloquial term meaning scattering from extraneous / spurious sources). The sample to detector distance is 6\,m on MAPS and is evacuated to $\sim 10^{-6}$\,mbar to prevent icing of unshielded cryogenic equipment. The detectors themselves are not in vacuum, but are behind thin (1\,mm thick) aluminium windows. They are 1\,m long tubes arranged in modules to form a square array covering scattering angles of $\pm 20^{\circ}$ in the horizontal and vertical directions. There is then a narrow strip of four modules covering horizontal scattering angles from $20^{\circ}$ to $60^{\circ}$.

\section{Upgrade description}\label{s:upgrade}

\begin{figure*}[t]
\includegraphics*[scale=0.65,angle=0]{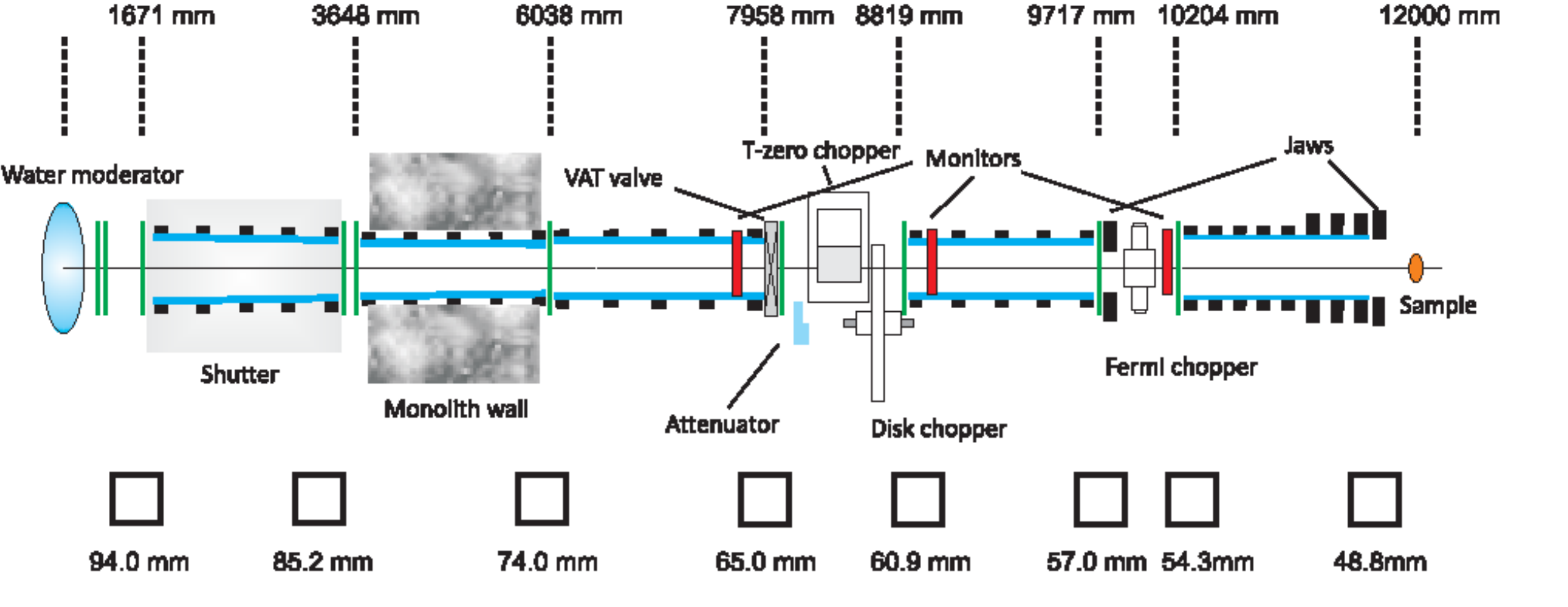}
\centering \caption{Schematic cross-section of the MAPS primary spectrometer after the upgrade (beam direction left to right). As in figure \ref{fig:layout_collimation} distances from the moderator and aperture sizes of components are given. Note that the first two monitors are mounted inside the vacuum housing of the guide, whereas previously these were outside the vacuum housing. Blue tapered lines indicate schematically the location of the glass guide substrate. Black squares indicate B$_{4}$C collars that both clamp the glass in place and also provide neutronic shielding.} \label{fig:layout_guide}
\end{figure*}

A schematic of the beamline layout after the upgrade is shown in figure \ref{fig:layout_guide}, and the key parameters are listed in table \ref{t:Inst_pars}. In short, the upgrade comprises replacement of all of the collimation with supermirror guides; addition of a disk chopper; and a modification to the poisoning of the ambient water moderator (details in sec. \ref{ss:moderator}) viewed by MAPS.

\begin{table}[!h]
\centering
\begin{tabular}{L{5cm} P{3.5cm}}
\hline\hline
\footnotesize{Moderator type} & \footnotesize{Ambient water, asymmetrically poisoned}\\
\footnotesize{Moderator -- T-zero chopper distance} & \footnotesize{8540\,mm}\\
\footnotesize{Moderator -- disk chopper distance} & \footnotesize{8831\,mm}\\
\footnotesize{Moderator -- Fermi chopper distance} & \footnotesize{10143\,mm}\\
\footnotesize{Fermi chopper -- sample distance} & \footnotesize{1857\,mm}\\
\footnotesize{Sample -- detector distance} & \footnotesize{6000\,mm}\\
\footnotesize{Guide m-value} & \footnotesize{3}\\
\footnotesize{Moderator -- monitor 1 distance} & \footnotesize{7933\,mm}\\
\footnotesize{Moderator -- monitor 2 distance} & \footnotesize{9327\,mm}\\
\footnotesize{Moderator -- monitor 3 distance} & \footnotesize{10326\,mm}\\
\footnotesize{Moderator -- monitor 4 distance} & \footnotesize{20392\,mm}\\
\footnotesize{Max. beam size at sample position} & \footnotesize{48 $\times$ 48\,mm}\\
\footnotesize{Max. Fermi chopper speed} & \footnotesize{600\,Hz}\\
\hline\hline
\end{tabular}
\caption{MAPS instrument parameters post-upgrade. Refer also to fig. \ref{fig:layout_guide} and text.}\label{t:Inst_pars}
\end{table}

\subsection{Guide}\label{ss:guide}

All of the sections of collimation were replaced by Ni/Ti supermirror guide with $m=3$, where $m$ is the critical angle of reflection of neutrons from a thick layer of natural nickel. The guide follows a straight taper from the entrance of the shutter to the sample position. It is notable that the aperture at the upstream end is larger than before (94\,mm square c.f. 73.3\,mm), providing a wider view of the moderator face, tapering to a slightly smaller beam size at the sample position (48.8\,mm c.f. 55.4\,mm). The guide follows a similar design to that employed on the MERLIN spectrometer at ISIS \cite{Bewley20061029} and was chosen mainly due to the constraints of space, i.e. a primary flight path of 12\,m and the need to re-use the existing T-zero chopper and Fermi chopper, which have fixed apertures. The aperture at the entrance and exit of each guide section is given in figure \ref{fig:layout_guide}. As with the collimation that it replaced, each guide section is terminated by a thin (1\,mm thick) aluminium vacuum window. The guide sections from the insert to the Fermi chopper share a common vacuum system, as with the old collimation in these sections. The guide in the shutter is also evacuated, rather than purged with helium as the old shutter collimation was. The final section of guide shares a common vacuum with the detector tank and hence has no window at the downstream end, eliminating the possibility of spurions from neutrons scattered by a window near the sample position. The guide substrate was float glass for the sections in the shutter and insert, and borofloat glass for the others. It is well-known \cite{BOFFY201614} that borofloat glass can be damaged in high thermal neutron radiation environments such as that found near to the source in the shutter and insert guide sections, which is why float glass was used for the guide substrate in these sections.

To minimize background several steps were taken. The glass substrate for the guide was kept to a minimal thickness of 3.5\,mm, largely to avoid the possibility of diffuse scattering of high energy neutrons. The vacuum vessel was therefore the heavy steel shielding surrounding the guide, and this was stepped on each guide section to avoid sight lines for fast neutrons from the target station to the sample. The guide was held in place and aligned using sintered B$_{4}$C collars, which also provide additional neutronic shielding around the guide. Each guide section had a sintered B$_{4}$C mask on each end that matched the beam aperture. Finally, as with the collimation that it replaced, the shielding for the final guide section was designed to avoid a line of sight to the walls of the guide housing. This was achieved by using similar sintered B$_{4}$C collars as the other guide sections, but with a decreasing spacing between them approaching the sample position. The final few collars were also deeper than those preceding, for the same reason.

The change in flux and beam divergence as a result of installing a supermirror guide was modelled using the McStas ray tracing software package \cite{Mcstas-paper}. As mentioned above, the choice of geometry was highly constrained by the need to re-use as many of the old beamline components as possible. The choices available were therefore what $m$-value of guide to use, and whether to follow a continuous straight taper of the guide from the source to the sample or whether to allow straight sections or even use an elliptical guide. An elliptical guide was ruled out due to the difficulty and lack of gain over a more conventional geometry when used on a short 12\,m primary flight path. An elliptical geometry would also not have allowed the choppers with their existing apertures to be re-used. More detailed consideration was given to whether to have any guide sections that were not tapered. Considered individually such guide sections would have an advantage over a tapered guide, with very similar flux gains (in fact higher flux gains at higher neutron energies at which the guide's critical angle is small) and slightly reduced divergence. However the constraint of getting the beam through the existing chopper apertures meant that using a non-tapered guide section on one part of the beamline would require a steeper taper on another guide section downstream, negating the benefits described above.

\begin{figure}[!h]
\includegraphics*[scale=0.6,angle=0]{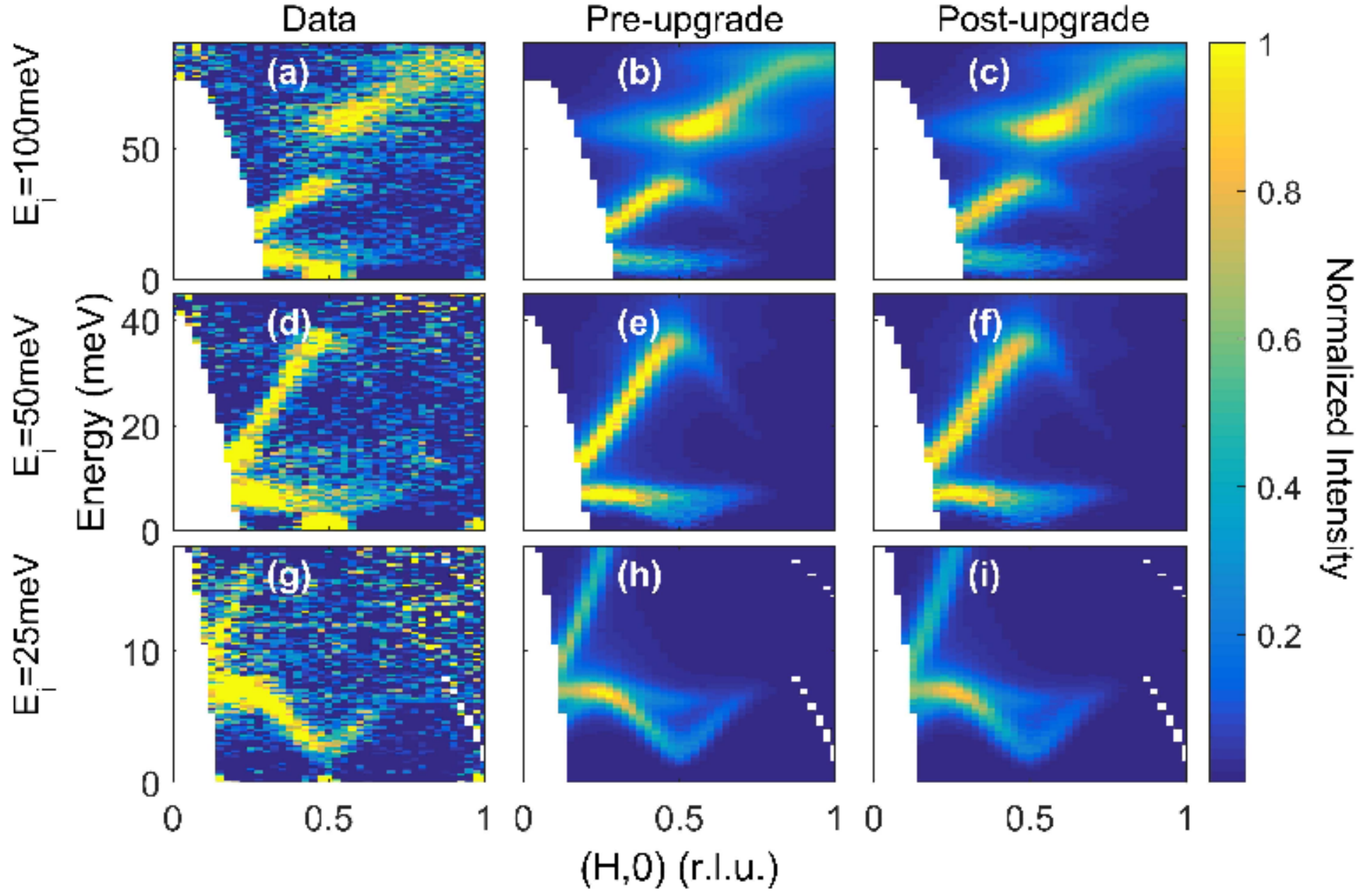}
\centering \caption{Illustration of the expected change in resolution before and after the upgrade. Panels (a), (d) and (g) show data collected by Johnston {\it et al} on Pr(Ca,Sr)$_{2}$Mn$_{2}$O$_{7}$ using MAPS before the upgrades were installed \cite{Johnstone-PRL-12} with incident energies $E_{i}=100, 50, 25$\,meV respectively. Panels (b), (e) and (h) show the corresponding resolution-convoluted simulation of the cross section for the pre-upgrade instrument. Panels (c), (f) and (i) show the expected resolution-convoluted simulation of the cross section for the post-upgrade instrument with $m=3$ guide.} \label{fig:pcsmo}
\end{figure}

Concerning the choice of guide $m$-value, beam divergence was a key consideration. The long (6\,m) secondary flight path of MAPS gives high resolution in both energy transfer and in momentum transfer, $\mathbf{Q}$, compared to shorter instruments such as MERLIN. A qualitative decision had to be taken regarding how much extra divergence, and hence degraded $\mathbf{Q}$ resolution, would be acceptable. This was done by simulating the beam divergence for a range of neutron energies for several guide $m$-values, as well as the changed moderator pulse width resulting from the change in the poisoning (see sec. \ref{ss:moderator}). These numbers were then used with the Tobyfit software \cite{Tobyfit-ref} to do a resolution-convoluted simulation of the `known' scattering cross-section from analyses that were previously performed \cite{Johnstone-PRL-12} on MAPS on a sample of Pr(Ca,Sr)$_{2}$Mn$_{2}$O$_{7}$. These simulations were then compared with the results obtained from the old MAPS instrument (i.e. with no guide). A comparison between the data (panels (a), (d) and (g)), the simulated cross-section with no guide (panels (b), (e) and (h)) and the simulated cross-section for the proposed upgrade with an $m=3$ guide (panels (c), (f) and (i)) is shown in figure \ref{fig:pcsmo} for reference. It can be seen that with $E_{i}=100$\,meV and $E_{i}=50$\,meV there is no qualitative difference between pre- and post-upgrade. For $E_{i}=25$\,meV there is a slight broadening visible in the signal arising from the dispersive spin waves, however the effect is minor. This gave confidence that the increased divergence resulting from the upgrade would not have a detrimental effect on the resolution for typical experiments.

\begin{figure}[!h]
\includegraphics*[scale=0.6,angle=0]{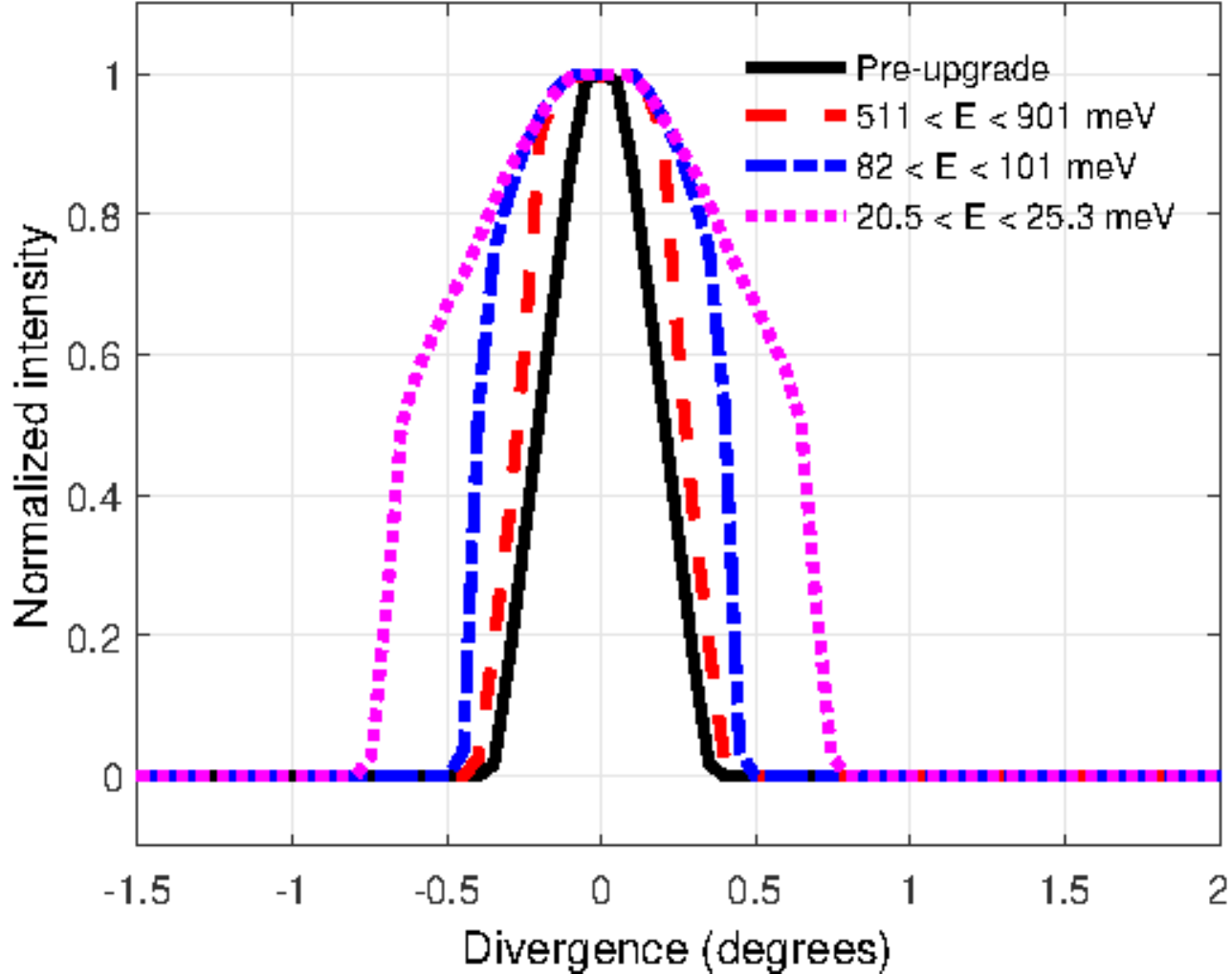}
\centering \caption{Simulated beam divergence at the sample position as a function of incident energy $E_{i}$ for the pre-upgrade instrument, and for an $m=3$ straight tapered guide. The black solid line shows the mean divergence of the beam before the upgrade, which is set entirely by the instrument geometry and is independent of $E_{i}$. The red dashed line shows the mean divergence for energies in the range $511 < E < 901$\,meV, the blue dash-dot line shows the mean divergence for energies in the range $82 < E < 101$\,meV and the magenta dotted line shows the divergence for energies in the range $20.5 < E < 25.3$\,meV. Note that these energy ranges correspond to neutron wavelength intervals of $0.3 < \lambda < 0.4$\,$\textrm{\AA}$, $0.9 < \lambda < 1.0$\,$\textrm{\AA}$ and $1.8 < \lambda < 2.0$\,$\textrm{\AA}$ respectively.} \label{fig:divergence}
\end{figure}

The simulated mean divergence for three representative energies with an $m = 3$ guide are shown in fig. \ref{fig:divergence}, with the four curves showing the results for $511 < E_{i} < 909$\,meV, $82 < E_{i} < 101$\,meV and $20.5 < E_{i} < 25.3$\,meV with the guide, and the divergence for the instrument without a guide. Pre-upgrade the divergence is set entirely by the collimation geometry, and hence is energy-independent. The simulated gain in flux is shown in fig. \ref{fig:ratiolam1} and is discussed in section \ref{ss:flux}.

\begin{figure}[!h]
\includegraphics*[scale=0.65,angle=0]{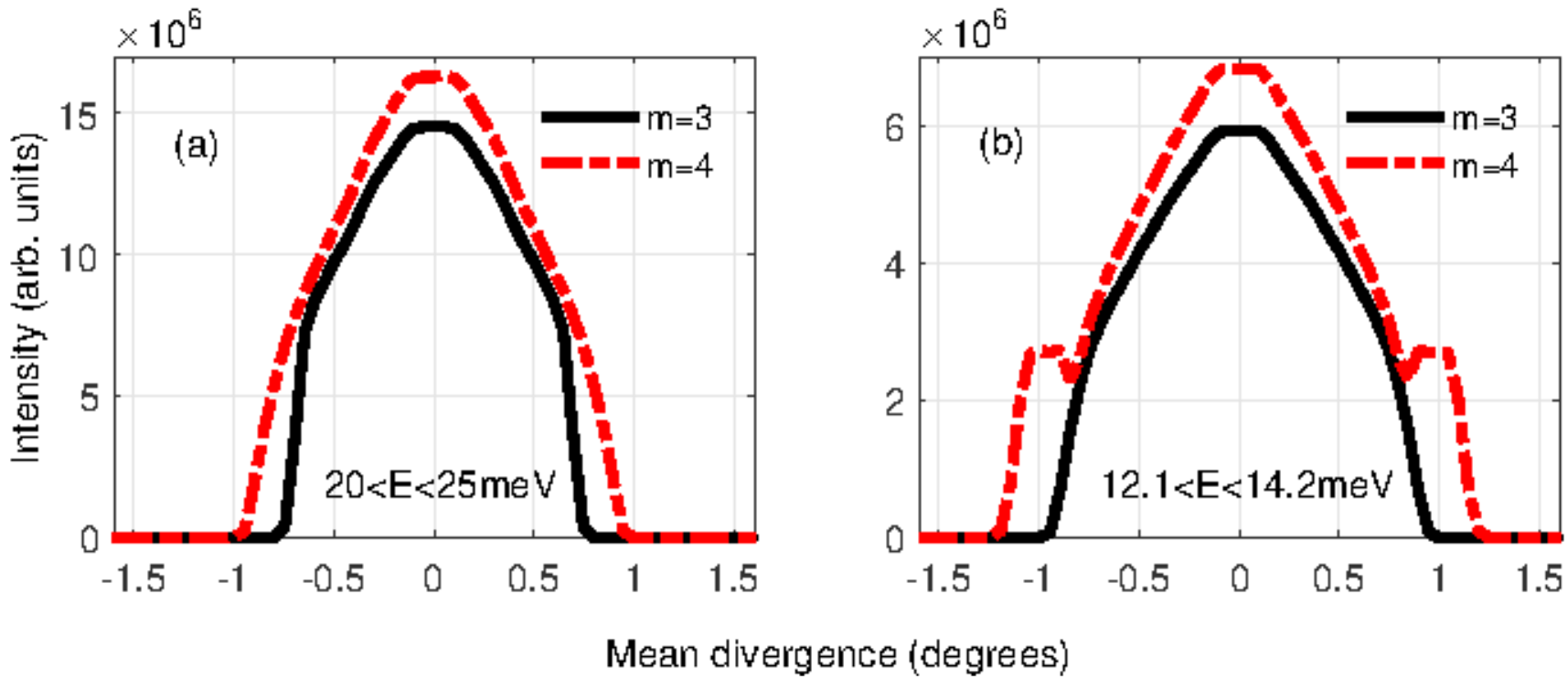}
\centering \caption{Simulated beam divergence at the sample position for an $m=3$ and an $m=4$ guide (black solid line and red dashed line respectively). Panel (a) shows the case for neutrons in the energy range $20 < E < 25$\,meV; panel (b) shows the case for neutrons in the energy range $12.1 < E < 14.2$\,meV.} \label{fig:guide_div_comp}
\end{figure}

We performed the same resolution-convoluted simulation of the cross-section as shown in fig. \ref{fig:pcsmo} for an $m=4$ guide, and found that there was little qualitative worsening of the resolution for any of the incident energies shown. However there were other reasons we chose not to use a guide with $m > 3$. An important consideration is the divergence profile at different energies. In fig. \ref{fig:guide_div_comp} we illustrate a problem with an $m=4$ guide at the low energy end of the normal operating range of MAPS. For $20 < E < 25$\,meV the divergence is slightly larger for $m=4$ than for $m=3$, however as noted above the qualitative effect on data quality is likely to be rather small. However for $12.1 < E < 14.2$\,meV the divergence profile for $m=4$ is much more structured than for $m=3$. This would, for $m=4$, lead to unusual resolution effects that would be very challenging to model and would have a significant negative impact on data quality. Another key consideration is what happens to very low energy neutrons which are able to pass through the chopper system when it is phased for a higher $E_{i}$, i.e. a kind of spurion. In table \ref{t:S-Emin} we list the minimum energy that can be transmitted by the MAPS `sloppy' Fermi chopper for several different frequencies of operation. We found that for $m=4$ such spurions have a divergence large enough that the direct beam is incident on the lowest angle detectors for Fermi chopper frequencies $\leq 200$\,Hz, whereas for $m=3$ this situation only arises for frequencies $\leq 100$\,Hz. We note that in the last eight years of operation the Fermi chopper has only been used at 100\,Hz for $0.3\%$ of measurements, and has never been used at 50\,Hz. On the other hand the instrument was regularly operated with the Fermi chopper running at 150\,Hz or 200\,Hz.

We note that even for an $m=3$ guide neutrons with energies $\lesssim 1.7$\,meV have sufficient divergence that a portion of the direct beam impinges on the lowest angle detectors if there is no chopper system in place to stop them. Traditionally detector diagnostics and calibration are performed by measuring the scattering from a vanadium sample with an incident white beam. However any detectors on which the direct beam impinges cannot be calibrated in this way, and must therefore either be masked or calibrated using a different method. This has operational consequences for the instrument team, but should not impact on users who are employing the Fermi chopper system (i.e. the overwhelming majority) as described above.

\subsection{Moderator}\label{ss:moderator}

During the time when the upgrade was being conceived an historical analysis was performed of calibration measurements performed on MAPS since its inception \cite{Skoro-moderators}. Because the same vanadium plate sample had always been used to measure in white beam mode (in which the Fermi chopper is removed, used to check detector performance and perform standard calibrations) and in monochromatic mode (with the Fermi chopper, used for absolute flux normalization) there was a 15 year record of the instrument's flux profile and typical modes of operation respectively. Two key findings from this analysis were that, (a) the measured flux had been declining very slowly over a number of years, with the figure in 2015 about 10\% lower than it had been in the early 2000s; (b) for about 90\% of experiments the choice of Fermi chopper slit pack and frequency were such that the contribution to the energy resolution of the chopper term was substantially larger than the moderator term, indicating that the moderator pulse width was too narrow in most cases.

\begin{figure}[ht]
\includegraphics*[scale=0.6,angle=0]{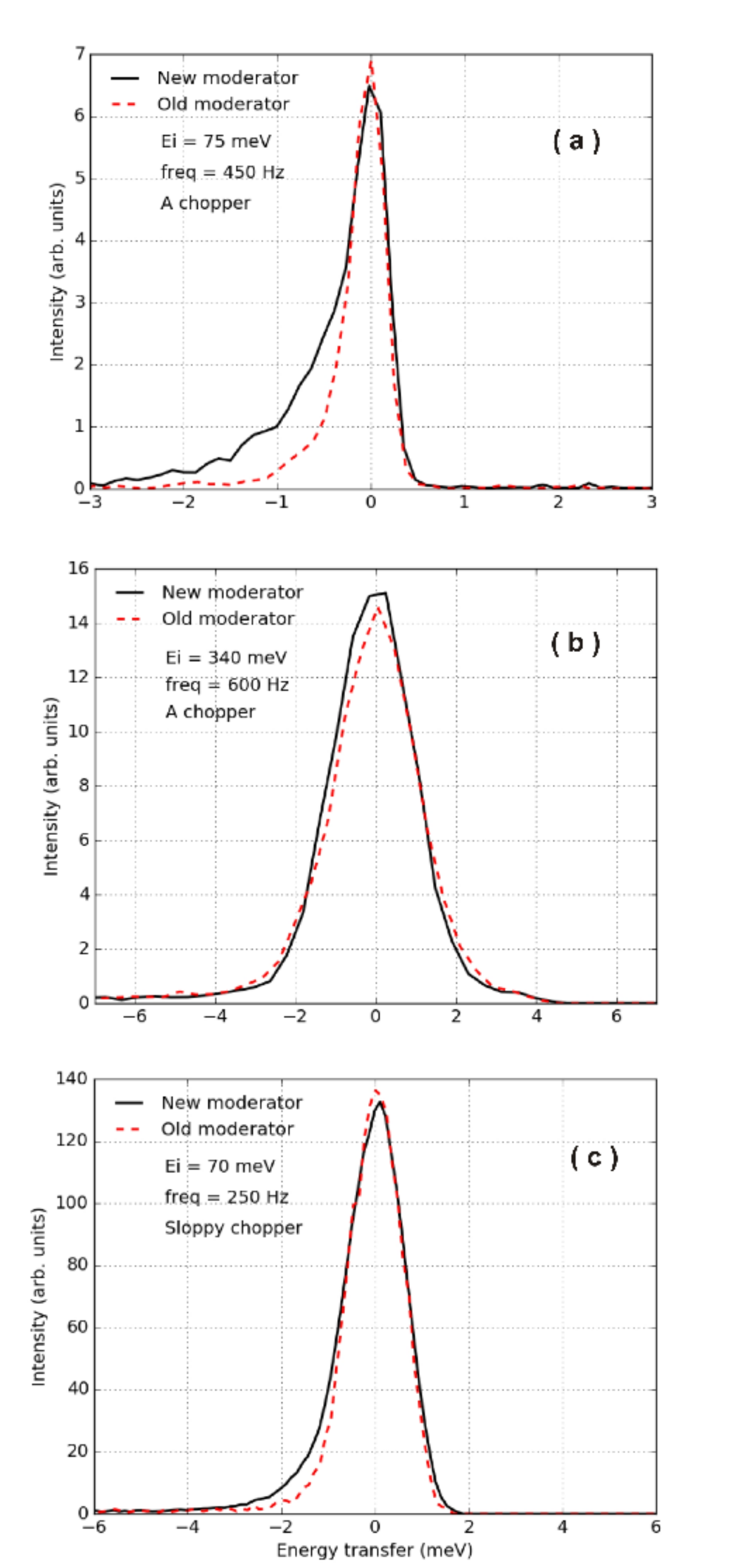}
\centering \caption{Typical time-of-flight spectra before (red dashed lines) and after (black solid lines) the moderator upgrade. Data have been scaled to the same peak height so that pulse widths can be more easily compared. (a) Incident energy of 75\,meV with the high-resolution `A' chopper running at 450\,Hz. In this condition the moderator contribution to the pulse width is dominant. (b) Incident energy of 350\,meV with the `A' chopper running at 600\,Hz. Because this energy is higher than the Gd cut-off no difference is expected in the pulse width before and after the moderator upgrade. (c) Incident energy of 70\,meV with the lower-resolution 'sloppy' chopper running at 250\,Hz. This is a typical configuration in which the instrument is run, where the chopper contribution to the resolution dominates over the moderator term.} \label{fig:pulses}
\end{figure}

The original design of the decoupled ambient temperature water moderator viewed by MAPS comprised an aluminium vessel divided into three roughly equal volumes by two gadolinium poisoning foils parallel to the moderator face. Neutronic calculations \cite{Skoro-moderators} indicated that if the poisoning foil closest to the moderator face viewed by MAPS was removed then an increase in flux of a factor $\sim 2$ would be expected. Although this would come at a cost of broadened pulse width, because the chopper term was dominating the resolution of MAPS in most cases anyway this would actually have only a minor effect on the energy resolution. A moderator following this design was installed early in 2016 and the measured increase in flux was found to be as expected. A comparison of the moderator pulse shape for three incident energies and chopper settings is shown in fig. \ref{fig:pulses}, which shows the TOF spectra measured at the detectors. In panel (a) a case is shown in which the high-resolution `A' chopper is used for a rather low incident energy, for which the moderator term in the resolution is expected to be relatively large. As expected, the increased moderator pulse width is visible in the form of a broadened asymmetric lineshape. In panel (b) a case for an incident energy above the gadolinium cutoff energy is shown. Here the change in poisoning is expected to have no effect, since at these neutron energies the gadolinium is almost transparent anyway. In panel (c) a representative pulse is shown using the low-resolution `sloppy' chopper rotating at a frequency typical of standard operation. Here there is a very marginal broadening of the instrumental resolution, but this penalty is acceptable when one considers the flux in this condition is twice as high.

\subsection{Choppers}\label{ss:choppers}

\begin{figure}[!h]
\includegraphics*[scale=0.3,angle=0]{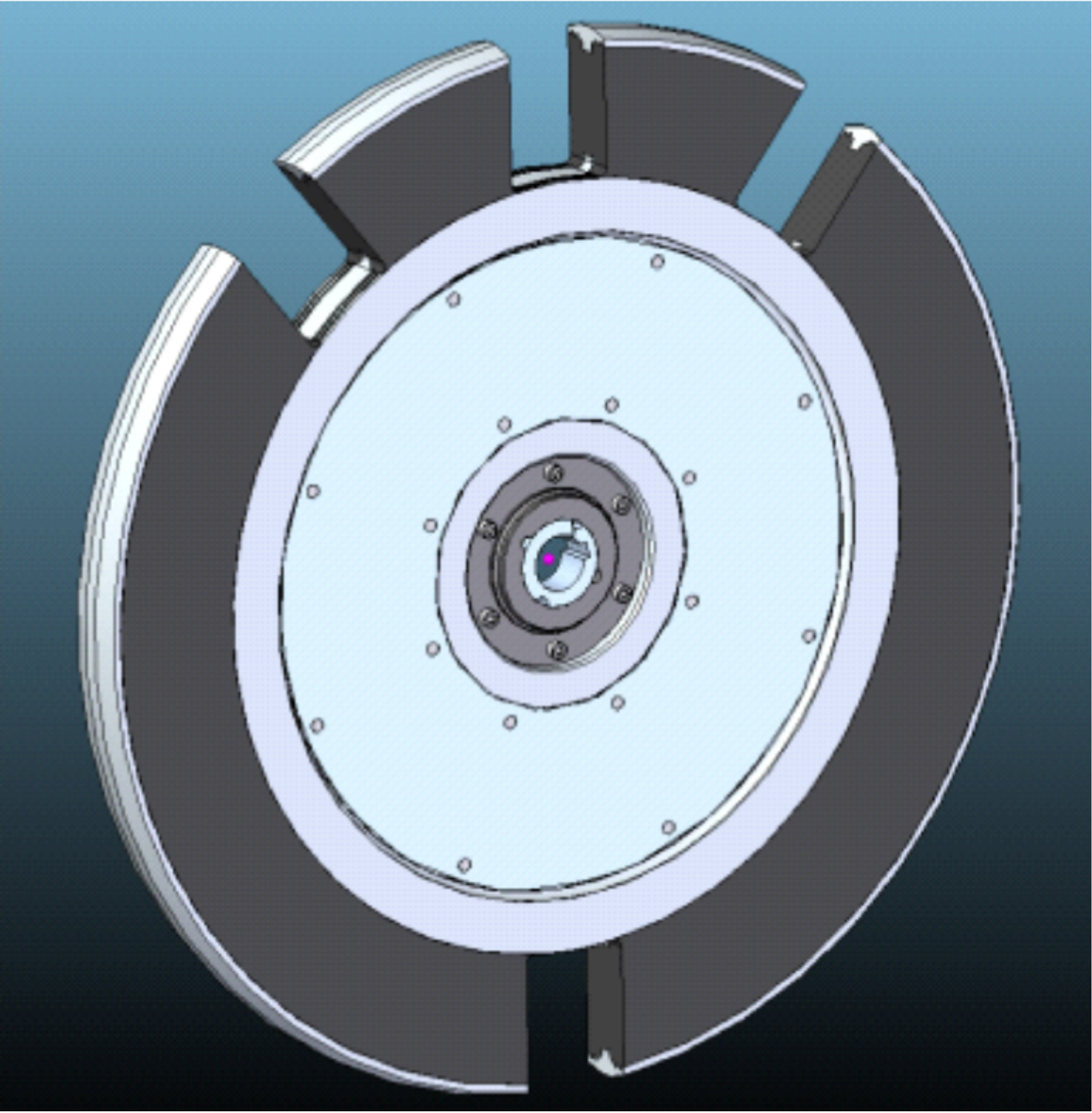}
\centering \caption{Disk chopper slot layout. The three slots at the top of the disk are for the RRM option and are spaced to allow transmission of multiple pulses through the Fermi chopper when it is running at 400\,Hz and the disk is operating at the ISIS source frequency of 50\,Hz. The single slot on the bottom of the disk is for when the Fermi chopper is running at other frequencies. It is used to suppress the background from the so-called `$\pi$-pulse' and neutron energy gain scattering therefrom.} \label{fig:disk}
\end{figure}

As well as retaining the T-zero and Fermi choppers from the old instrument, the upgraded instrument also employs a disk chopper, situated between the T-zero and Fermi choppers at 8.831\,m from the source. A schematic of the disk is shown in fig. \ref{fig:disk}. There are a total of four slots in the disk, three equally spaced square slots whose widths and heights are defined by the straight line between the upstream and downstream tapered guide sections' apertures, and a fourth slot on the opposite side of the disk the same width and height as the other three. The purpose of this disk chopper is two-fold.

\begin{figure}[!h]
\includegraphics*[scale=0.42,angle=0]{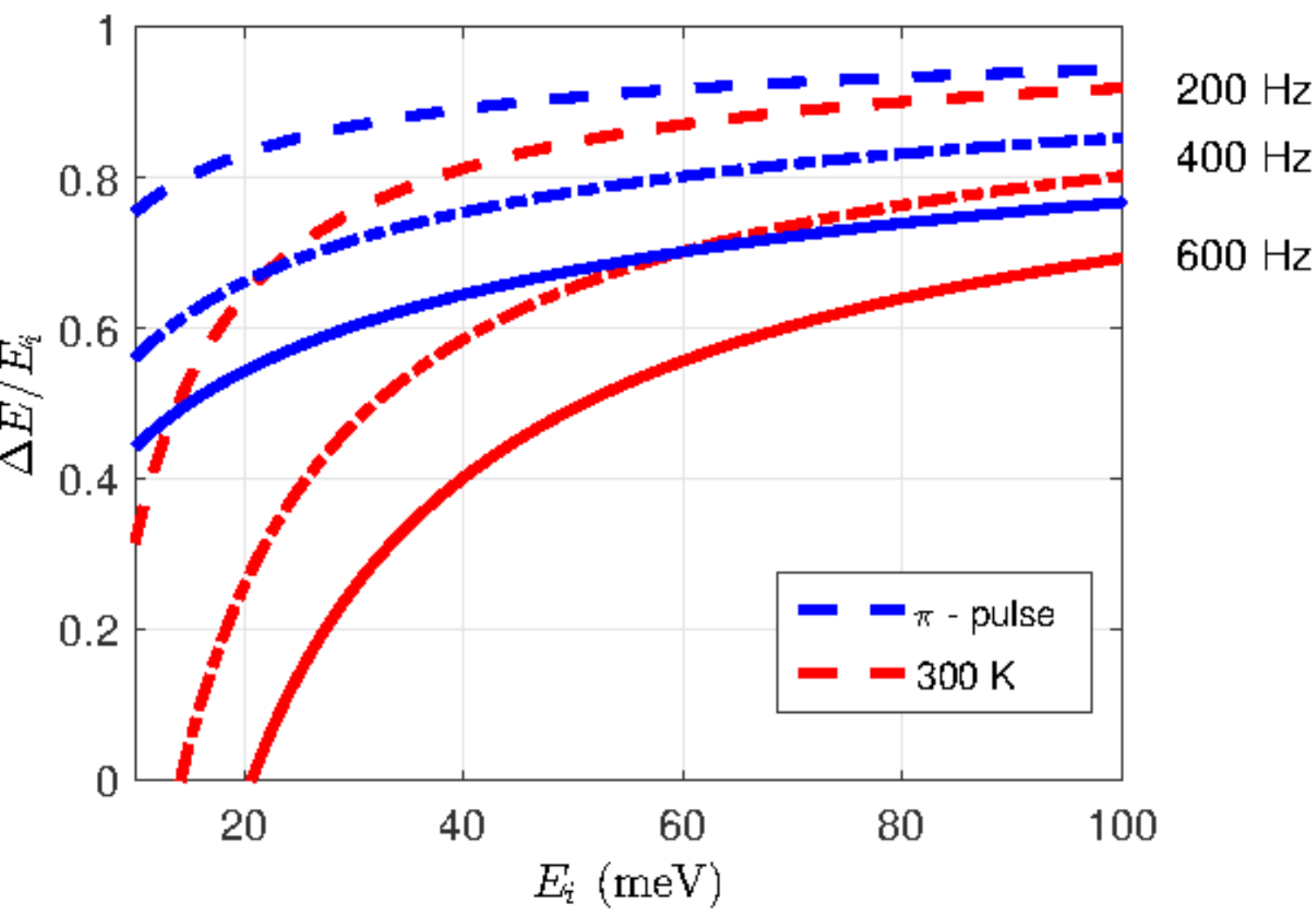}
\centering \caption{Blue lines: apparent location of the $\pi$-pulse in energy transfer as a fraction of $E_{i}$ for a given $E_{i}$ for three different choices of Fermi chopper frequency ($\nu$). The dashed line is for the case of $\nu = 200$\,Hz, the dash-dot line is for $\nu = 400$\,Hz, and the solid line is for $\nu = 600$\,Hz. Red lines, following the same line style for each Fermi chopper frequency, indicate the apparent location in energy transfer as a fraction of $E_{i}$ of neutron energy gain scattering of 30\,meV by the sample when $\pi$-pulse neutrons are incident on it. Energy gain of 30\,meV was chosen to correspond approximately to a sizeable thermal population factor at room temperature $\sim 300$\,K.} \label{fig:chop_spurions}
\end{figure}

When the disk is phased so that the single slot allows through a range of neutron energies centered on the energy to be selected by the Fermi chopper its role is background suppression. By ensuring that the upstream line of sight is shut off at earlier and later openings of the Fermi chopper the background is reduced. This is most noticeable for the next opening of the Fermi chopper after the pulse of interest (the so-called $\pi$-pulse), and particularly for the case when samples are being measured at elevated temperatures. In this case there would be substantial neutron energy gain scattering of the $\pi$-pulse neutrons that impinges onto the frame of interest. This situation is illustrated in fig. \ref{fig:chop_spurions}. Here we show where in the frame of interest (in energy transfer as a fraction of $E_{i}$) the $\pi$-pulse arrives, and also where neutrons which have undergone energy gain scattering from a room-temperature sample, as a function of $E_{i}$ and Fermi chopper frequency. This shows that for lower $E_{i}$ and/or high Fermi chopper frequencies (which would be more common on MAPS after the substantial increases in flux afforded by the guide) these spurions can impinge on a substantial part of the region of interest. Eliminating the $\pi$-pulse with the disk chopper thus improves the background, especially for samples that are measured at elevated temperatures, and allows energy transfers that are a greater fraction of $E_{i}$ to be measured.

\begin{figure}[!h]
\includegraphics*[scale=0.45,angle=0]{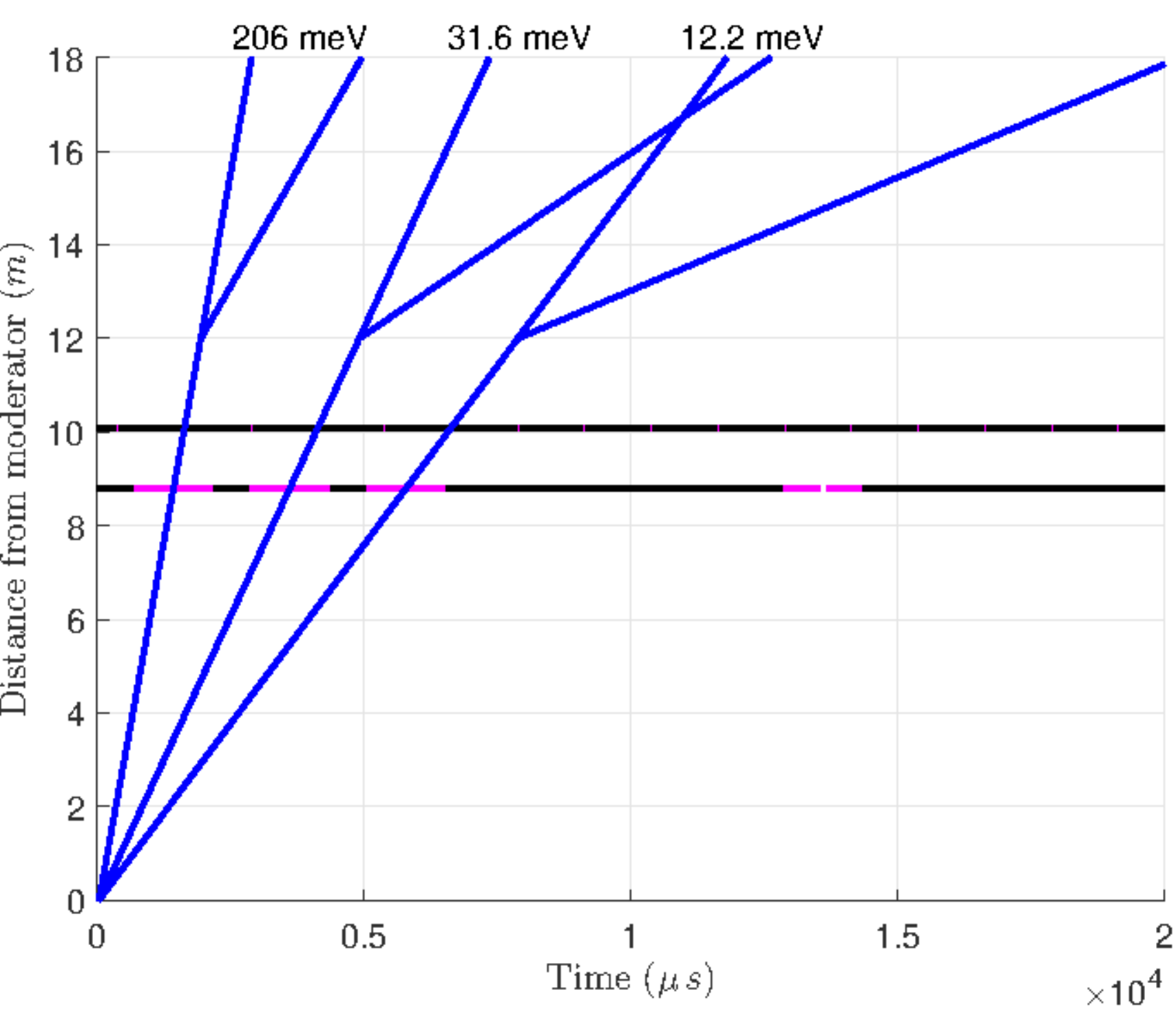}
\centering \caption{Repetition rate multiplication time-distance diagram for 400\,Hz Fermi chopper operation. The disk chopper (running at 50\,Hz) is located 8.83\,m from the source, the Fermi chopper 10.14\,m from the source, and the sample 12\,m from the source. Rays after the sample are for elastic scattering and for inelastic energy loss scattering with energy transfer of 90\% of $E_{i}$.} \label{fig:multirep}
\end{figure}

The second mode of operation of the disk chopper uses the three slits that are closer together. The spacing of the slits was chosen to ensure that when the Fermi chopper is run at 400\,Hz or 200\,Hz the line of sight to the source is open at the right time to allow multiple neutron energies through the Fermi chopper - a mode of operation known as "repetition rate multiplication" (RRM). This is now routinely employed on a number of instruments and is an effective way of allowing measurements at multiple energies to be performed in parallel, thus minimizing the amount of unused neutron detection time between source pulses \cite{Russina-RRM,Bewley-LET,CuGeO3_multirep}. Figure \ref{fig:multirep} illustrates the paths of neutrons through the instrument for a typical setup at 400\,Hz. In order to allow neutrons with the right trajectory for 400\,Hz operation of the Fermi chopper to be transmitted while the disk chopper is running at 50\,Hz the slots have an angular spacing of $\theta_{\rm{slot}}$ given by,

\begin{equation}\label{eq:fslot}
\theta_{\rm{slot}} = \frac{d_{\rm{Disk}}}{d_{\rm{Fermi}}} \times \frac{\nu_{\rm{Disk}}}{\nu_{\rm{Fermi}}} \times 360^{\circ},
\end{equation}

\noindent where $d_{\rm{Disk}}$ is the distance of the disk chopper from the source, $d_{\rm{Fermi}}$ is the distance of the Fermi chopper from the source, $\nu_{\rm{Disk}}$ is the frequency at which the disk chopper runs, and $\nu_{\rm{Fermi}}$ is the frequency at which the Fermi chopper runs. In the case for which the MAPS chopper system was designed this means that the angular separation of slot centers is $\theta_{\rm{slot}} = \frac{1}{9.18} \times 360^{\circ} = 39.21^{\circ}$.

We decided to use three slots whose spacing is essentially hard-wired for just two Fermi chopper frequencies, rather than have a single large slot that allows greater flexibility, as has been done on MERLIN. One reason for this is the relatively long secondary flight path on MAPS, 6\,m as opposed to 2.5\,m on MERLIN for example. Attempting to use RRM for Fermi chopper frequencies any faster than 400\,Hz results in significant `frame overlap', i.e. less and less of each pulse is useful before neutrons from the next pulse transmitted by the chopper arrive at the detectors (see above discussion of the $\pi$-pulse). Another disadvantage of a single large slot is that when the instrument is being run in single energy mode it means that there is a greater chance of low energy neutrons going through the large slot and then being transmitted by the Fermi chopper and arriving at the detector at some unexpected time. The minimum neutron energy that can be transmitted by the MAPS `sloppy' chopper as a function of frequency was calculated using {\sc Pychop}\cite{Pychop-ref}, part of the Mantid suite of programs\cite{Arnold2014_Mantid,Mantid-website}, and the results are given in table \ref{t:S-Emin}. If the Fermi chopper is running at high speed this is less of an issue because these low energy neutrons will not be transmitted, but it could be a serious problem if the Fermi chopper speed is lower or if a new slit pack was made that was optimized for the transmission of low energy neutrons. For example, if the large slot was open for neutrons with TOF in the range $8000 < \rm{TOF} < 12000 \mu s$ then all neutrons with energies in the range $2.81 \leq E \leq 6.33$\,meV would arrive at the Fermi chopper and could potentially be transmitted and arrive at the detectors in the next ISIS frame, i.e. after the next source pulse. Of course this would also have been true for the old instrument, with no disk chopper, however there the flux of neutrons with these kinds of energies was vanishingly small, whereas with the guide their flux is a factor $\sim 50$ higher. As already noted in sec. \ref{ss:guide}, with a guide a crucial point is that the divergence of these neutrons is very large so the direct beam may impinge on the low angle detectors.

\begin{table}[!h]
\centering
\begin{tabular}{P{3.5cm} P{5cm}}
\hline\hline
\footnotesize{Chopper frequency (Hz)} & \footnotesize{Minimum transmitted energy (meV)}\\
\hline
\footnotesize{100} & \footnotesize{0.82}\\
\footnotesize{150} & \footnotesize{1.9}\\
\footnotesize{200} & \footnotesize{3.3}\\
\footnotesize{250} & \footnotesize{5.1}\\
\footnotesize{300} & \footnotesize{7.4}\\
\footnotesize{350} & \footnotesize{10.0}\\
\footnotesize{400} & \footnotesize{13.0}\\
\hline\hline
\end{tabular}
\caption{Minimum energy transmitted by the MAPS `sloppy' chopper as a function of frequency.}\label{t:S-Emin}
\end{table}


\section{Commissioning results}\label{s:commissioning}

\subsection{Flux}\label{ss:flux}

\begin{figure}[!h]
\includegraphics*[scale=0.55,angle=0]{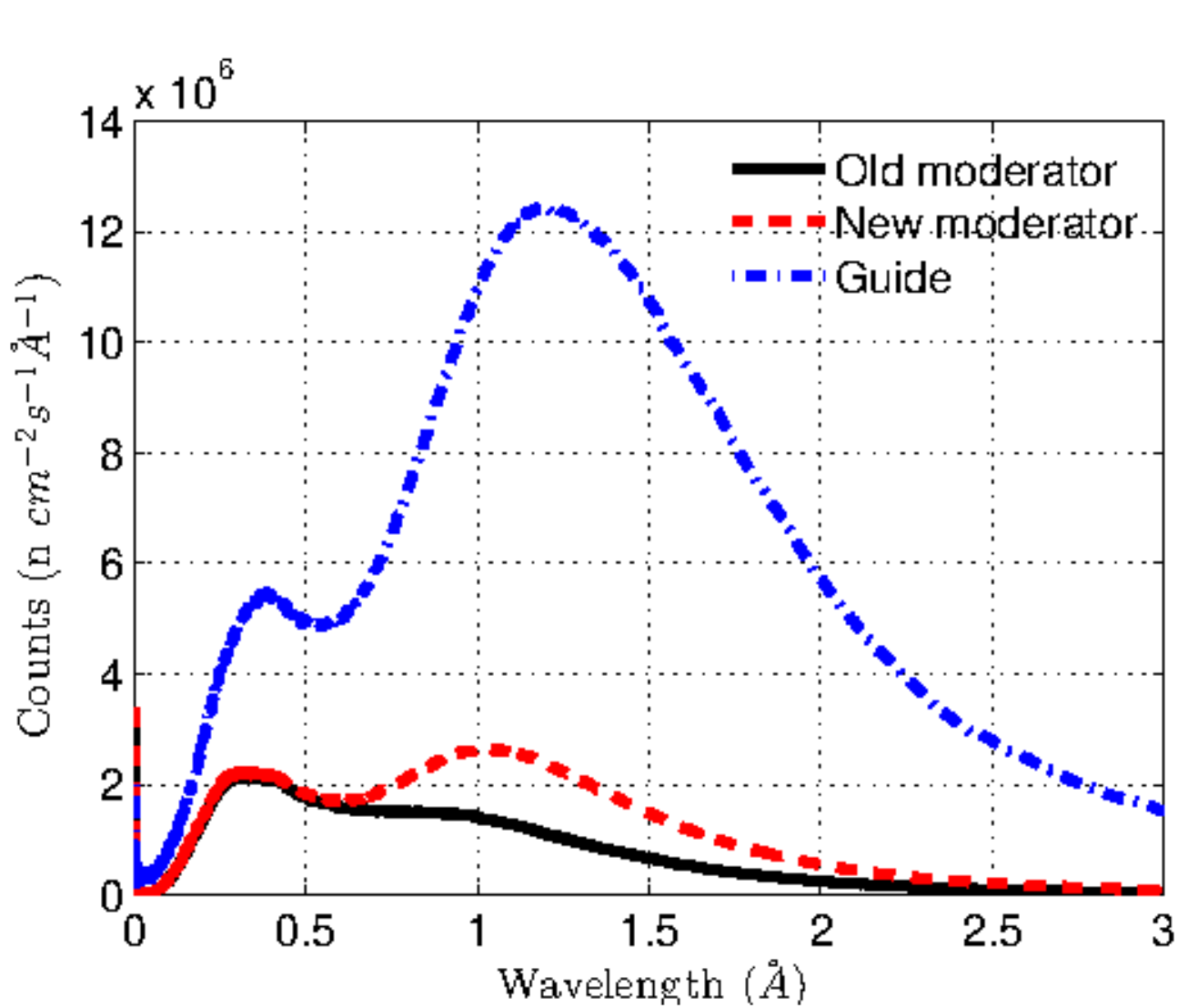}
\centering \caption{Measured MAPS white beam flux as a function of neutron wavelength for before the upgrade (black solid line), after the moderator change (red dashed line) and after the guide installation (blue dot-dashed line). Data collection method described in the text.} \label{fig:fluxlam1}
\end{figure}

In figure \ref{fig:fluxlam1} we show the white beam flux measured on MAPS in three conditions -- before the upgrade, after the change of moderator, and after the installation of the guide. The measurements were performed using the same flat-plate vanadium calibration sample scattering into the MAPS $^{3}$He detectors, summing the response of the detectors in the equatorial plane with scattering angles in the range $10^{\circ} \leq 2 \theta \leq 14^{\circ}$. As expected, significant gains in flux are seen with the installation of the guide, including at the very highest energies due to the increased beam aperture. In the peak flux condition for wavelengths around $1.2 \textrm{\AA}$ the total gain factor is an order of magnitude.

\begin{figure}[!h]
\includegraphics*[scale=0.4,angle=0]{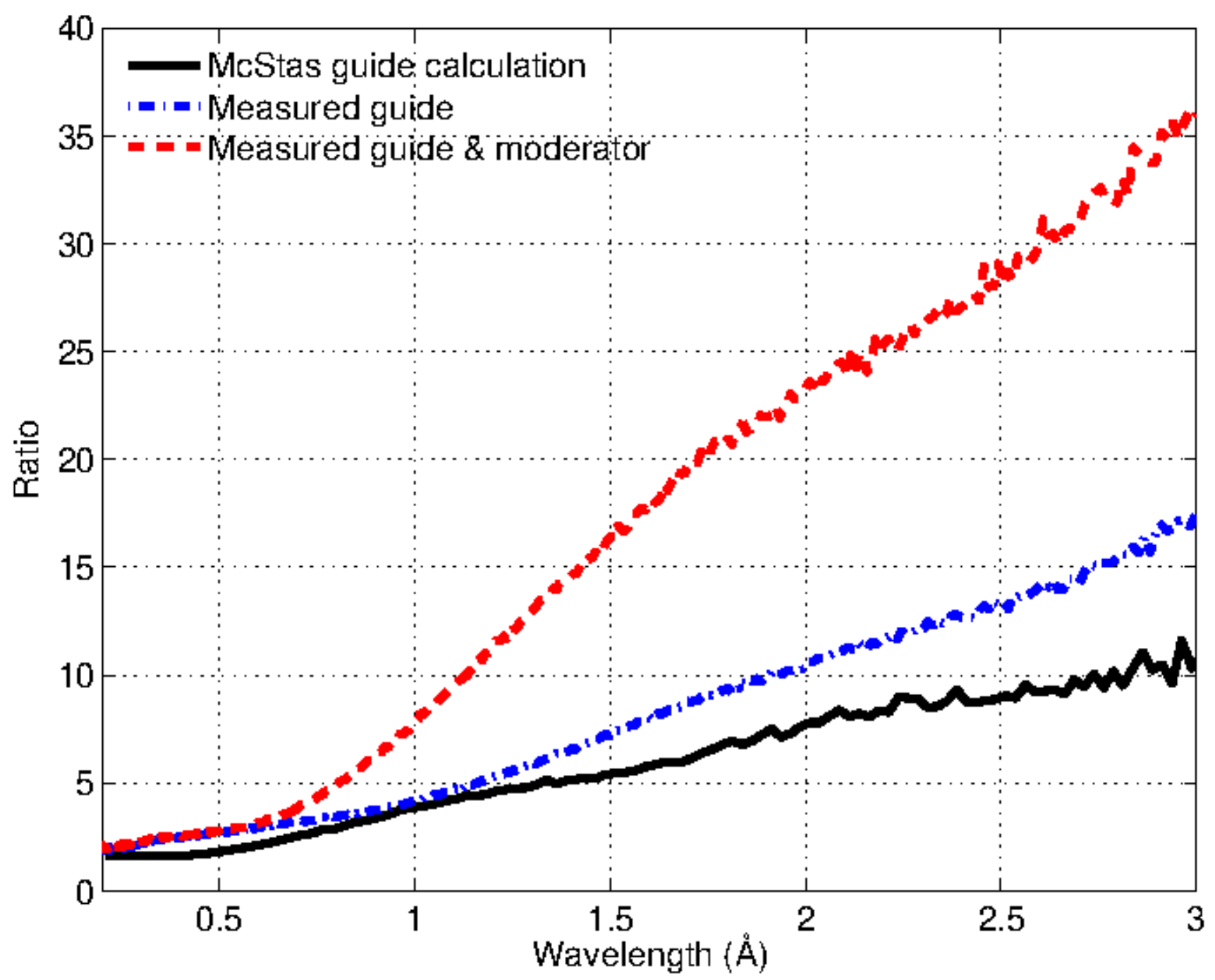}
\centering \caption{Measured and calculated gain factors from the guide and moderator as a function of neutron wavelength. The black solid line shows the McStas-simulated gain factor from the guide, the blue dot-dashed line shows the gain factor from the guide that was actually measured, and the red dashed line shows the measured gain including the moderator and the guide upgrades together.} \label{fig:ratiolam1}
\end{figure}

In figure \ref{fig:ratiolam1} we show the flux gain factor for the guide, the guide plus moderator, and a McStas simulation of the gain from installing a guide. It is notable that the gain factor measured is in excess of that predicted from the McStas simulations at all wavelengths. However this result is not surprising when placed in the context of problems that occurred on MAPS before the guide was installed \cite{MAPS-FluxLoss}, in which an unexplained issue with the old shutter caused a dramatic reduction in flux. To remedy this, the old shutter was replaced with the new one (containing neutron supermirror guides) before the rest of the guide was installed, and then the measured increase in flux was found to be $\sim 35\%$ higher than expected from McStas simulations. This suggests that the performance of the old instrument was below what one would expect from the engineering drawings, giving rise to the larger than expected gain observed when it was replaced.

\begin{figure}[!h]
\includegraphics*[scale=0.7,angle=0]{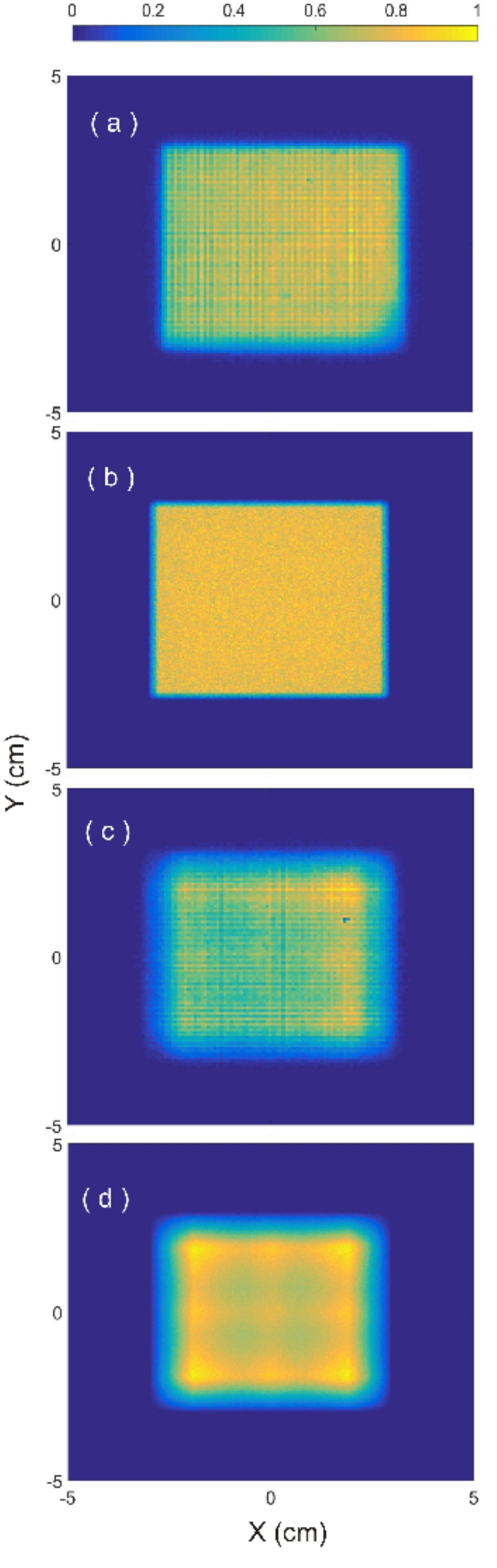}
\centering \caption{Colormaps of the flux (normalized to unity) at the MAPS sample position. Panels (a) and (b) show, respectively, the measured and McStas-simulated beam profile at the sample position for neutrons in the wavelength range $2 < \lambda < 2.3 \textrm{\AA}$ before the guide upgrade. Panels (c) and (d) show, respectively, the measured and simulated beam profile for the same wavelength range after the guide upgrade.} \label{fig:profile}
\end{figure}

With no guide and just collimation one would expect the beam profile at the sample position to be almost uniform, whereas a more patterned structure would be expected after the guide was installed. To validate this we performed TOF-resolved beam profile measurements using a neutron gas electron multiplier (nGEM) area detector \cite{ngem-ref}. This is a 2D neutron detector with an active area of $100 \times 100$\,mm$^{2}$ and a pixel size of $0.8 \times 0.8$\,mm$^{2}$. It uses a 100\,nm layer of $^{10}$B$_{4}$C as a neutron converter providing an efficiency of $10^{-4}$ for neutrons with a wavelength of 1\,$\textrm{\AA}$. It is able to operate up to a total count rate of 10\,MHz on the whole detector area. The data collected for neutrons with wavelength in the range $2 < \lambda < 2.3 \textrm{\AA}$ before and after the guide installation are shown in figure \ref{fig:profile}, together with McStas simulations of the beam profile for these wavelengths. At these wavelengths the guide is expected to have a noticeable effect on the beam profile, and indeed this is observed, with an inhomogeneity of around $\pm 20\%$. There is also a much less sharp cut-off in intensity around the edge of the beam after the guide installation, again in line with simulations.

\subsection{Resolution}\label{ss:resolution}

The installation of a guide is expected to have a negligible effect on the energy (TOF) resolution of the instrument. We verified this by performing a series of measurements using the vanadium standard sample described above with various $E_{i}$, chopper frequencies and chopper slit packs before and after the guide installation, similar to those shown in fig. \ref{fig:pulses}. No statistically significant effect was observed. The effect on the wavevector resolution is discussed in section \ref{ss:science}. This is expected to be more significant (see fig. \ref{fig:divergence}) because of the increased beam divergence, especially for incident energies below $\sim 100$\,meV.

\subsection{Scientific examples}\label{ss:science}

\begin{figure}[!h]
\includegraphics*[scale=0.5,angle=0]{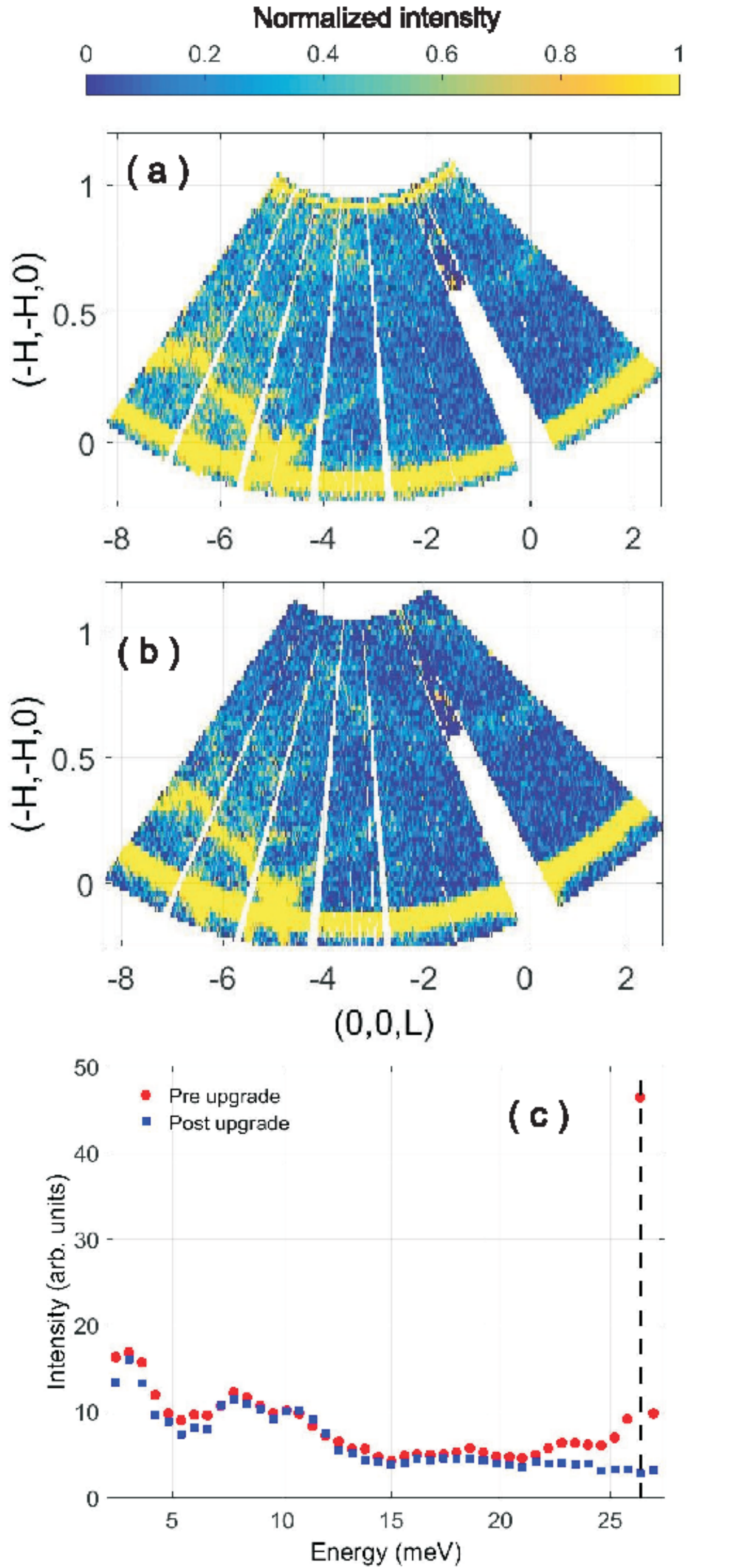}
\centering \caption{Projection on to the $(-H,-H,L)$-plane of a measurement of La$_{1.84}$Sr$_{0.16}$CuO$_{4}$ at room temperature. The data were taken in a single sample orientation (detailed method described in the text) and show an acoustic phonon. Panel (a) shows data collected before the guide and moderator upgrade in a time of 280 minutes, and panel (b) shows data collected after the upgrade in a time of 20 minutes. Both datasets were taken with $E_{i}=30.1$\,meV with the `sloppy' chopper spinning at 200\,Hz and the post-upgrade data were taken with the disk chopper set to single energy mode. Panel (c) shows both datasets integrated over the full Q-range as a function of energy transfer, and illustrates quantitatively the reduction in background above $\sim 21$\,meV. The vertical dashed line indicates the position of the $\pi$-pulse. Note that errorbars are smaller than the markers.} \label{fig:lsco}
\end{figure}

La$_{2-x}$Sr$_{x}$CuO$_{4}$ with $x=0.16$ is a cuprate high-temperature superconductor with $T_{c}=$38.5\,K. Samples of this series with various values of $x$ have been extensively measured on MAPS in the past \cite{Vignolle_NPhys07,Headings_PRL10,Lipscombe-PRL-09,Lipscombe-PRL07}, providing a useful benchmark for the instrument's performance. The group which published refs. \onlinecite{Vignolle_NPhys07,Headings_PRL10,Lipscombe-PRL-09,Lipscombe-PRL07} has used measurements of an acoustic phonon as a way of cross-checking the absolute units calibration from the vanadium standard in all of its measurements, thus providing a useful historical record of the instrument's flux and resolution. Re-measuring this phonon provides three tests. First, it allows a comparison of counting times before and after the upgrade that are required to obtain data of similar statistical quality. Second it allows us to verify the effect of the guide and moderator upgrade on the resolution in a typical instrument configuration. Third, because the measurement is performed at room temperature it allows us to illustrate the background reduction effect of the disk chopper. Data were collected with $E_{i}=30.1$\,meV with the `sloppy' chopper spinning at 200\,Hz and the disk chopper set in single energy mode. The sample (of mass 35.4\,g) was oriented in the $(H,H,0) / (0,0,L)$ scattering plane and measurements were taken with the sample orientation\footnote{$\psi = 0$ refers to the case when the incident beam is parallel to the $(1,1,0)$-direction} $\psi = 65$. The data shown in panels (a) and (b) of figure \ref{fig:lsco} are a projection of the curved 3-dimensional hypersurface in 4-dimensional $\mathbf{Q}$-energy space, where one of the $\mathbf{Q}$-axes is coupled to energy (approx. radial in the segment shown in the figure) and the signal is integrated over a finite range of another $\mathbf{Q}$-axis to produce a 2-dimensional map of the scattering. The intense scattering near the bottom of the images is the incoherent elastic line and the acoustic phonon can be seen dispersing from $\mathbf{Q}=(0,0,-5)$. There is no qualitative degradation of the resolution when comparing panels (a) and (b). The data taken before the upgrade were measured over 280\,minutes (panel (a)) whereas the data taken after the upgrade (panel (b)) had similar statistical quality after 20\,minutes, in line with what would be expected from the flux gain at this energy. The background reduction from the disk chopper is clearly visible, particularly at higher energy transfers (larger values of $(-H,-H,0)$) and most notably with the removal of the intense band of scattering near $(-1,-1,L)$ which corresponds to the $\pi$-pulse. To illustrate this effect better we show a cut in panel (c) along the energy axis where the $\mathbf{Q}$ integration is $-8 \leq L \leq -2$, $-0.2 \leq (-H,-H,0) \leq 1.2$ and $-0.05 \leq (H,-H,0) \leq 0.05$. Above about 20\,meV a clear upturn is visible in the pre-upgrade cut compared to that taken post-upgrade, which corresponds to the energy gain scattering from the $\pi$-pulse, and the $\pi$-pulse itself at about 26\,meV is completely suppressed in the post-upgrade scan. This means that weaker features at high energies can be discerned in the post-upgrade data which would never have been visible before the upgrade.

\begin{figure}[h]
\includegraphics*[scale=0.5,angle=0]{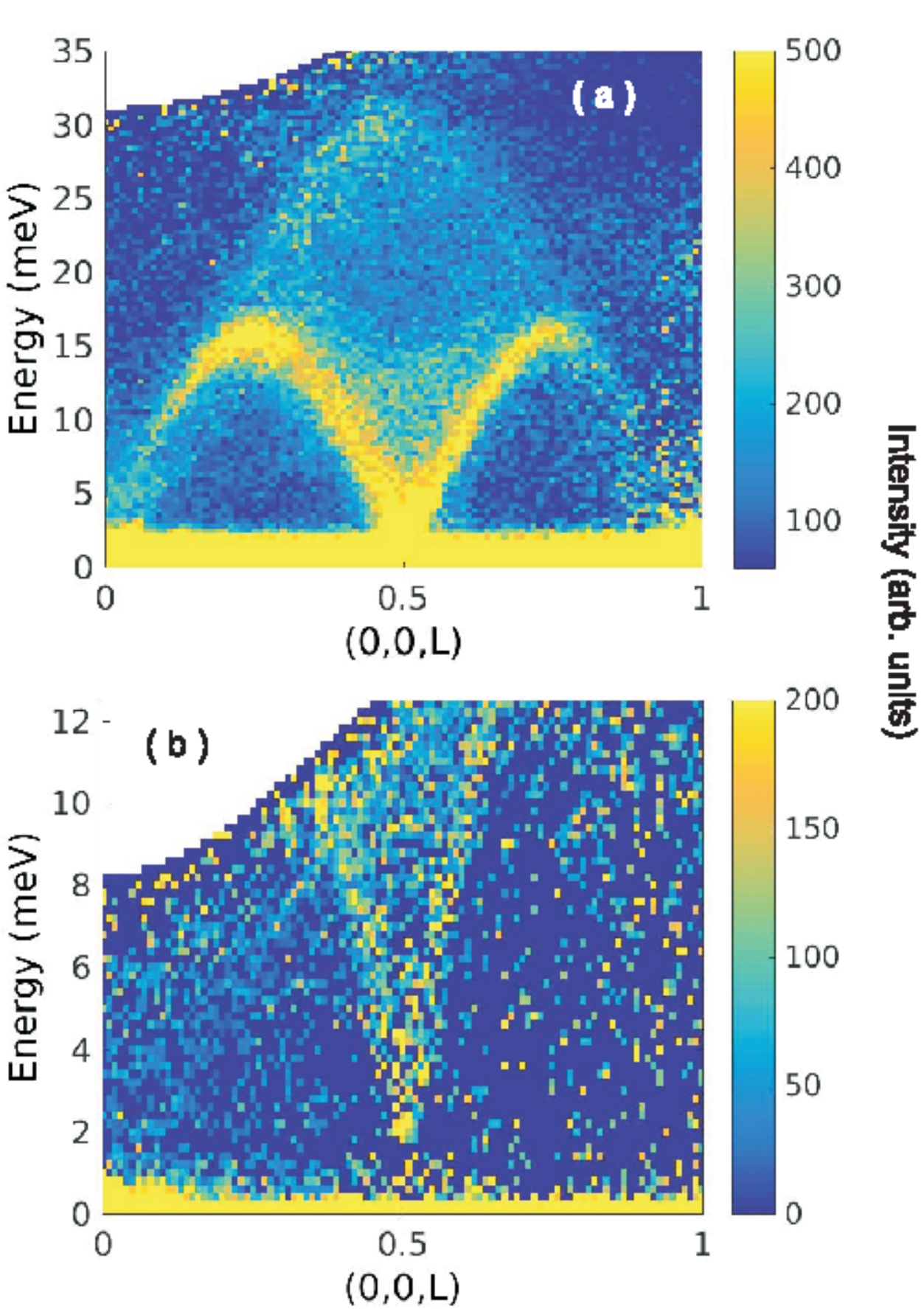}
\centering \caption{The spinon continuum in the one-dimensional $S=1/2$ chain compound CuGeO3. The data were collected in multirep mode with a Horace scan in the range $0 \leq \psi \leq 90$. Panel (a) shows data collected with $E_{i}=70$\,meV and panel (b) shows data collected with $E_{i}=19.5$\,meV.} \label{fig:cugeo3}
\end{figure}

CuGeO$_{3}$ is a one-dimensional $S=1/2$ spin chain compound that has been extremely well-studied using inelastic neutron scattering. Indeed, it has been used to gauge the performance of numerous neutron spectrometers at various stages of their development \cite{CuGeO3-4seasons,CuGeO3_multirep,Arai-CuGeO3-overview,Kajimoto-CuGeO3,Kajimoto-NIMA}. This makes it an attractive prospect to illustrate the characteristics of the upgraded MAPS spectrometer. A single crystal sample of mass $\sim 3$\,g was cooled to 5\,K. Data were collected with an incident energy of 70\,meV with the `sloppy' chopper spinning at 400\,Hz, thus allowing use of the RRM option and giving a second $E_{i}$ of 19.5\,meV. A so-called `Horace scan' \cite{Ewings-Horace} was performed in which the sample was rotated through $90^{\circ}$ in $2^{\circ}$ steps in the $bc$-plane, starting from $(0,0,1)$ parallel to the incident beam. A measurement of 15 minutes duration was performed for each sample orientation, giving a total measurement time of 11.5 hours. Data were then integrated along the directions orthogonal to the chain axis (the $c$-axis) in the ranges $-1 \leq H \leq 1$ and $-2 \leq K \leq 2$.  The data are shown in figure \ref{fig:cugeo3}. The expected two-spinon continuum is clearly visible for the higher $E_{i}$ dataset as a broad area of diffuse scattering up to 30\,meV with sharp boundaries with a periodicity of one reciprocal lattice unit (r.l.u.) for the upper bound and 0.5 r.l.u. for the lower bound, at which there is significant additional intensity. The lower $E_{i}$ dataset provides a high resolution measurement of the gap of $\sim 2$\,meV in the lower boundary of the continuum at $L = 1/2$. We note that the statistics in the lower $E_{i}$ dataset are qualitatively worse than the for the higher $E_{i}$ dataset because in this case at 19.5\,meV the chopper pulse width is substantially smaller than the moderator pulse width, meaning that only a small fraction of the available neutrons are selected. Measurements of pseudo-one-dimensional spin chain compounds such as this have often in the past been performed using a single sample orientation, with the chain axis perpendicular to the incident beam and integration over the axes either explicitly or implicitly performed. The advantage of doing a Horace scan is that the integration axes are all known explicitly, which can help to ensure that extra contributions from the background are ameliorated. Also weak inter-chain dispersion can be probed, if present.

\begin{figure}[h]
\includegraphics*[scale=0.6,angle=0]{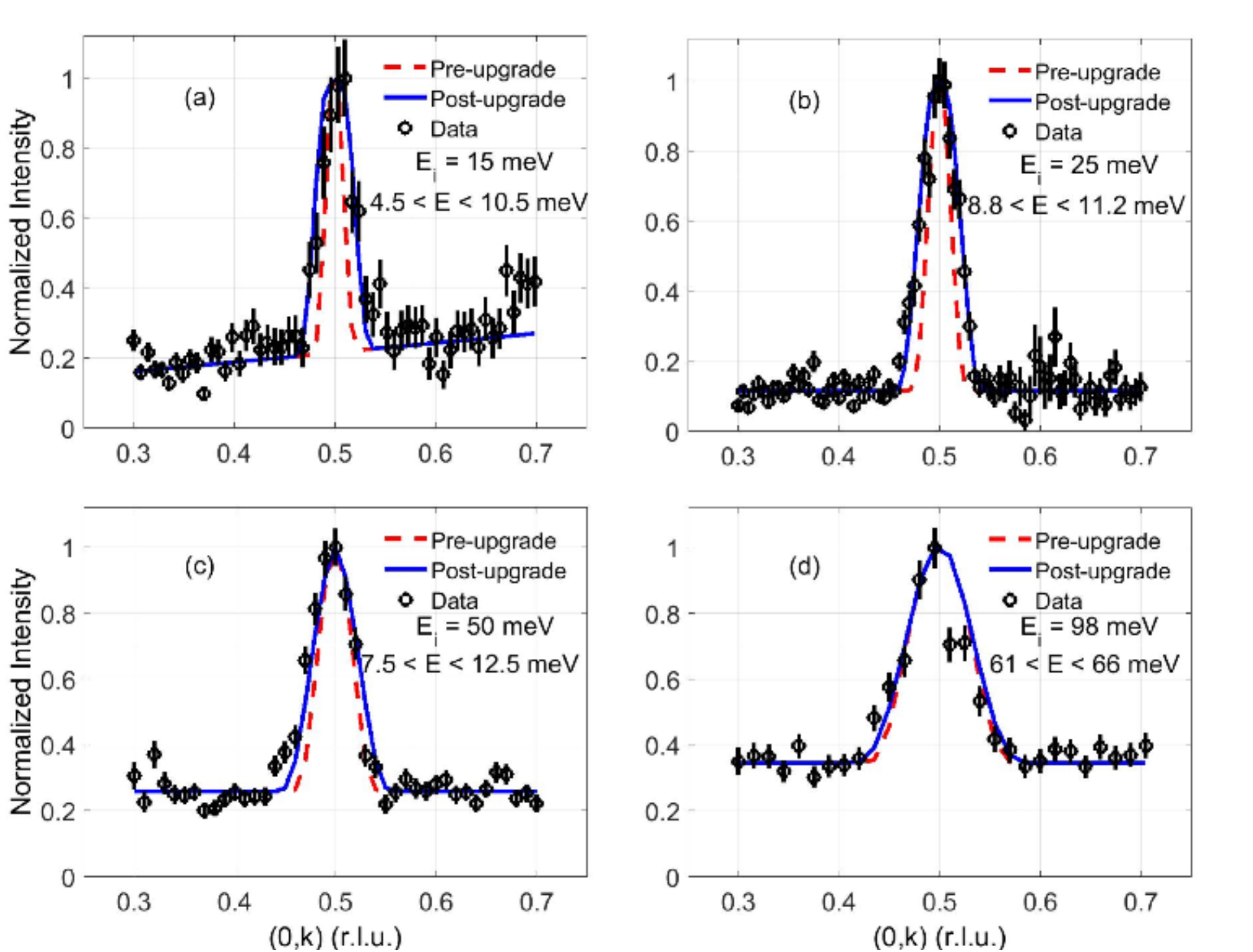}
\centering \caption{Data (black circles) taken from a sample of Sr$_{2}$CuO$_{3}$ after the instrument upgrade, together with simulations of the cross-section (see text) for the pre- and post-upgrade instrument (red dashed lines and blue solid lines respectively). Data are shown for (a) $E_{i}=15$\,meV and $4.5 \leq E \leq 10.5$\,meV; (b) $E_{i}=25$\,meV and $8.8 \leq E \leq 11.2$\,meV; (c) $E_{i}=50$\,meV and $7.5 \leq E \leq 12$\,meV; and (d) $E_{i}=98$\,meV and $61 \leq E \leq 66$\,meV. These energy transfer ranges were chosen to avoid background signals arising from phonons that were nearby in $(\mathbf{Q},E)$-space.} \label{fig:sco}
\end{figure}

Like CuGeO$_{3}$, Sr$_{2}$CuO$_{3}$ is a one-dimensional $S=1/2$ system that had been studied in some detail previously on MAPS \cite{Walters-NPhys-09}. It can be modelled exactly \cite{Caux-JStatMech}, and the data taken previously by Walters {\it et al} on MAPS were resolution-limited in wavevector at low energy transfers because the dispersion of the lower boundary of the two-spinon continuum in Sr$_{2}$CuO$_{3}$ is extremely steep. In figure \ref{fig:sco} we show data taken on MAPS after the upgrade for several different values of $E_{i}$ and energy transfer, together with resolution-convoluted simulations using the parameters from the old instrument and from the present instrument. The simulations illustrate, and are verified by the data, the noticeable degradation in Q-resolution for lower values of $E_{i}$, but only minor degradation at higher values of $E_{i}$. This can be understood with reference to fig. \ref{fig:divergence}, which shows that this energy-dependent effect arises entirely due to the increased angular divergence of the incident beam when transported by the guide. The resolution convolution for the simulations was performed using the {\sc Tobyfit} software \cite{Tobyfit-ref}, and for the post-upgrade instrument the increased beam divergence due to the guide was modelled using a enlarged effective moderator aperture, as had been used previously with success for data taken on the MERLIN spectrometer \cite{Babbers-Cu-cupolas,Ewings-Sr122}.

We noted in section \ref{s:intro} that approximately one quarter of the time on MAPS is used for molecular spectroscopy experiments. These experiments are usually complementary to inelastic neutron scattering measurements of the same samples using indirect geometry instruments such as the TOSCA spectrometer at ISIS \cite{Tosca-paper}. On instruments like TOSCA high quality data with good count rate and superior energy resolution to MAPS can be obtained for low energies, $\lesssim 150$\,meV. However, at higher energies only data at high $\mathbf{Q}$ are accessible, for which the damping of the signal by the Debye-Waller factor \cite{Tranquada-book} is significant \cite{Parker-review}. Because the kinematic constraints of a direct geometry instrument like MAPS are different, and allow access to much smaller $\mathbf{Q}$ at high energy transfers, molecular spectroscopy experiments using MAPS are focused on incident energies $E_{i}\gtrsim 200$\,meV. This in turn means that the increase in flux from the upgrade is around a factor 2, and arises solely from the increased solid angle of the moderator viewed by the instrument, in the region of interest for molecular spectroscopy experiments.


\section{Summary}\label{s:summary}

We have described a programme of upgrades to the MAPS time of flight chopper spectrometer. These comprise a change to the poisoning of the water moderator viewed by MAPS, installation of supermirror neutron guides to replace the collimation between the source and sample, and installation of a new disk chopper. The performance of all of the components is either in line with, or in excess of, simulations. The boost to the flux in the thermal and sub-thermal neutron energy range will significantly enhance the capabilities of MAPS.


\section{Acknowledgements}\label{s:ack}
We are very grateful to the technical teams at ISIS, led by John Randall and Paul Cross, for their dedication and hard work which ensured that the installations of all the upgrades proceeded in a timely fashion and were completed without any major difficulties. We also acknowledge the user groups who loaned us samples to help commission the instrument, in particular Igor Zaliznyak, Stephen Hayden, Lauren Cane, Stewart Parker, David Lennon and Andrew Princep.

\bibliography{refs_maps_paper}

\begin{thebibliography}{73}%
\makeatletter
\providecommand \@ifxundefined [1]{%
 \@ifx{#1\undefined}
}%
\providecommand \@ifnum [1]{%
 \ifnum #1\expandafter \@firstoftwo
 \else \expandafter \@secondoftwo
 \fi
}%
\providecommand \@ifx [1]{%
 \ifx #1\expandafter \@firstoftwo
 \else \expandafter \@secondoftwo
 \fi
}%
\providecommand \natexlab [1]{#1}%
\providecommand \enquote  [1]{``#1''}%
\providecommand \bibnamefont  [1]{#1}%
\providecommand \bibfnamefont [1]{#1}%
\providecommand \citenamefont [1]{#1}%
\providecommand \href@noop [0]{\@secondoftwo}%
\providecommand \href [0]{\begingroup \@sanitize@url \@href}%
\providecommand \@href[1]{\@@startlink{#1}\@@href}%
\providecommand \@@href[1]{\endgroup#1\@@endlink}%
\providecommand \@sanitize@url [0]{\catcode `\\12\catcode `\$12\catcode
  `\&12\catcode `\#12\catcode `\^12\catcode `\_12\catcode `\%12\relax}%
\providecommand \@@startlink[1]{}%
\providecommand \@@endlink[0]{}%
\providecommand \url  [0]{\begingroup\@sanitize@url \@url }%
\providecommand \@url [1]{\endgroup\@href {#1}{\urlprefix }}%
\providecommand \urlprefix  [0]{URL }%
\providecommand \Eprint [0]{\href }%
\providecommand \doibase [0]{http://dx.doi.org/}%
\providecommand \selectlanguage [0]{\@gobble}%
\providecommand \bibinfo  [0]{\@secondoftwo}%
\providecommand \bibfield  [0]{\@secondoftwo}%
\providecommand \translation [1]{[#1]}%
\providecommand \BibitemOpen [0]{}%
\providecommand \bibitemStop [0]{}%
\providecommand \bibitemNoStop [0]{.\EOS\space}%
\providecommand \EOS [0]{\spacefactor3000\relax}%
\providecommand \BibitemShut  [1]{\csname bibitem#1\endcsname}%
\let\auto@bib@innerbib\@empty
\bibitem [{\citenamefont {Perring}\ \emph {et~al.}(1994)\citenamefont
  {Perring}, \citenamefont {Taylor}, \citenamefont {Osborn}, \citenamefont
  {Paul}, \citenamefont {Boothroyd},\ and\ \citenamefont {Aeppli}}]{MAPS-tech}%
  \BibitemOpen
  \bibfield  {author} {\bibinfo {author} {\bibfnamefont {T.~G.}\ \bibnamefont
  {Perring}}, \bibinfo {author} {\bibfnamefont {A.~D.}\ \bibnamefont {Taylor}},
  \bibinfo {author} {\bibfnamefont {R.}~\bibnamefont {Osborn}}, \bibinfo
  {author} {\bibfnamefont {D.~M.}\ \bibnamefont {Paul}}, \bibinfo {author}
  {\bibfnamefont {A.~T.}\ \bibnamefont {Boothroyd}}, \ and\ \bibinfo {author}
  {\bibfnamefont {G.}~\bibnamefont {Aeppli}},\ }\href@noop {} {\bibfield
  {journal} {\bibinfo  {journal} {RAL Report}\ }\textbf {\bibinfo {volume}
  {94-025}},\ \bibinfo {pages} {60} (\bibinfo {year} {1994})}\BibitemShut
  {NoStop}%
\bibitem [{\citenamefont {Princep}\ \emph {et~al.}(2017)\citenamefont
  {Princep}, \citenamefont {Ewings}, \citenamefont {Ward}, \citenamefont
  {Toth}, \citenamefont {Dubs}, \citenamefont {Prabhakaran},\ and\
  \citenamefont {Boothroyd}}]{Princep-YIG}%
  \BibitemOpen
  \bibfield  {author} {\bibinfo {author} {\bibfnamefont {A.~J.}\ \bibnamefont
  {Princep}}, \bibinfo {author} {\bibfnamefont {R.~A.}\ \bibnamefont {Ewings}},
  \bibinfo {author} {\bibfnamefont {S.}~\bibnamefont {Ward}}, \bibinfo {author}
  {\bibfnamefont {S.}~\bibnamefont {Toth}}, \bibinfo {author} {\bibfnamefont
  {C.}~\bibnamefont {Dubs}}, \bibinfo {author} {\bibfnamefont {D.}~\bibnamefont
  {Prabhakaran}}, \ and\ \bibinfo {author} {\bibfnamefont {A.~T.}\ \bibnamefont
  {Boothroyd}},\ }\href@noop {} {\bibfield  {journal} {\bibinfo  {journal} {NPJ
  Quantum Materials}\ }\textbf {\bibinfo {volume} {2}} (\bibinfo {year}
  {2017})}\BibitemShut {NoStop}%
\bibitem [{\citenamefont {Stock}\ \emph {et~al.}(2016)\citenamefont {Stock},
  \citenamefont {Rodriguez}, \citenamefont {Lee}, \citenamefont {Green},
  \citenamefont {Demmel}, \citenamefont {Ewings}, \citenamefont {Fouquet},
  \citenamefont {Laver}, \citenamefont {Niedermayer}, \citenamefont {Su},
  \citenamefont {Nemkovski}, \citenamefont {Rodriguez-Rivera},\ and\
  \citenamefont {Cheong}}]{Stock-CFO}%
  \BibitemOpen
  \bibfield  {author} {\bibinfo {author} {\bibfnamefont {C.}~\bibnamefont
  {Stock}}, \bibinfo {author} {\bibfnamefont {E.~E.}\ \bibnamefont
  {Rodriguez}}, \bibinfo {author} {\bibfnamefont {N.}~\bibnamefont {Lee}},
  \bibinfo {author} {\bibfnamefont {M.~A.}\ \bibnamefont {Green}}, \bibinfo
  {author} {\bibfnamefont {F.}~\bibnamefont {Demmel}}, \bibinfo {author}
  {\bibfnamefont {R.~A.}\ \bibnamefont {Ewings}}, \bibinfo {author}
  {\bibfnamefont {P.}~\bibnamefont {Fouquet}}, \bibinfo {author} {\bibfnamefont
  {M.}~\bibnamefont {Laver}}, \bibinfo {author} {\bibfnamefont
  {C.}~\bibnamefont {Niedermayer}}, \bibinfo {author} {\bibfnamefont
  {Y.}~\bibnamefont {Su}}, \bibinfo {author} {\bibfnamefont {K.}~\bibnamefont
  {Nemkovski}}, \bibinfo {author} {\bibfnamefont {J.~A.}\ \bibnamefont
  {Rodriguez-Rivera}}, \ and\ \bibinfo {author} {\bibfnamefont {S.~W.}\
  \bibnamefont {Cheong}},\ }\href@noop {} {\bibfield  {journal} {\bibinfo
  {journal} {Phys. Rev. Lett.}\ }\textbf {\bibinfo {volume} {117}} (\bibinfo
  {year} {2016})}\BibitemShut {NoStop}%
\bibitem [{\citenamefont {Dalla~Piazza}\ \emph {et~al.}(2014)\citenamefont
  {Dalla~Piazza}, \citenamefont {Mourigal}, \citenamefont {Christensen},
  \citenamefont {Nilsen}, \citenamefont {Tregenna-Piggott}, \citenamefont
  {Perring}, \citenamefont {Enderle}, \citenamefont {McMorrow}, \citenamefont
  {Ivanov},\ and\ \citenamefont {R{\o}nnow}}]{Dalla-Piazza_NPhys15}%
  \BibitemOpen
  \bibfield  {author} {\bibinfo {author} {\bibfnamefont {B.}~\bibnamefont
  {Dalla~Piazza}}, \bibinfo {author} {\bibfnamefont {M.}~\bibnamefont
  {Mourigal}}, \bibinfo {author} {\bibfnamefont {N.~B.}\ \bibnamefont
  {Christensen}}, \bibinfo {author} {\bibfnamefont {G.~J.}\ \bibnamefont
  {Nilsen}}, \bibinfo {author} {\bibfnamefont {P.}~\bibnamefont
  {Tregenna-Piggott}}, \bibinfo {author} {\bibfnamefont {T.~G.}\ \bibnamefont
  {Perring}}, \bibinfo {author} {\bibfnamefont {M.}~\bibnamefont {Enderle}},
  \bibinfo {author} {\bibfnamefont {D.~F.}\ \bibnamefont {McMorrow}}, \bibinfo
  {author} {\bibfnamefont {D.~A.}\ \bibnamefont {Ivanov}}, \ and\ \bibinfo
  {author} {\bibfnamefont {H.~M.}\ \bibnamefont {R{\o}nnow}},\ }\href@noop {}
  {\bibfield  {journal} {\bibinfo  {journal} {Nat. Phys.}\ }\textbf {\bibinfo
  {volume} {11}},\ \bibinfo {pages} {62} (\bibinfo {year} {2014})}\BibitemShut
  {NoStop}%
\bibitem [{\citenamefont {Qureshi}\ \emph {et~al.}(2014)\citenamefont
  {Qureshi}, \citenamefont {Steffens}, \citenamefont {Lamago}, \citenamefont
  {Sidis}, \citenamefont {Sobolev}, \citenamefont {Ewings}, \citenamefont
  {Harnagea}, \citenamefont {Wurmehl}, \citenamefont {Buechner},\ and\
  \citenamefont {Braden}}]{Qureshi-LiFeAs}%
  \BibitemOpen
  \bibfield  {author} {\bibinfo {author} {\bibfnamefont {N.}~\bibnamefont
  {Qureshi}}, \bibinfo {author} {\bibfnamefont {P.}~\bibnamefont {Steffens}},
  \bibinfo {author} {\bibfnamefont {D.}~\bibnamefont {Lamago}}, \bibinfo
  {author} {\bibfnamefont {Y.}~\bibnamefont {Sidis}}, \bibinfo {author}
  {\bibfnamefont {O.}~\bibnamefont {Sobolev}}, \bibinfo {author} {\bibfnamefont
  {R.~A.}\ \bibnamefont {Ewings}}, \bibinfo {author} {\bibfnamefont
  {L.}~\bibnamefont {Harnagea}}, \bibinfo {author} {\bibfnamefont
  {S.}~\bibnamefont {Wurmehl}}, \bibinfo {author} {\bibfnamefont
  {B.}~\bibnamefont {Buechner}}, \ and\ \bibinfo {author} {\bibfnamefont
  {M.}~\bibnamefont {Braden}},\ }\href@noop {} {\bibfield  {journal} {\bibinfo
  {journal} {Phys. Rev. B}\ }\textbf {\bibinfo {volume} {90}} (\bibinfo {year}
  {2014})}\BibitemShut {NoStop}%
\bibitem [{\citenamefont {Stock}\ \emph {et~al.}(2014)\citenamefont {Stock},
  \citenamefont {Rodriguez}, \citenamefont {Sobolev}, \citenamefont
  {Rodriguez-Rivera}, \citenamefont {Ewings}, \citenamefont {Taylor},
  \citenamefont {Christianson},\ and\ \citenamefont {Green}}]{Stock-FeTe}%
  \BibitemOpen
  \bibfield  {author} {\bibinfo {author} {\bibfnamefont {C.}~\bibnamefont
  {Stock}}, \bibinfo {author} {\bibfnamefont {E.~E.}\ \bibnamefont
  {Rodriguez}}, \bibinfo {author} {\bibfnamefont {O.}~\bibnamefont {Sobolev}},
  \bibinfo {author} {\bibfnamefont {J.~A.}\ \bibnamefont {Rodriguez-Rivera}},
  \bibinfo {author} {\bibfnamefont {R.~A.}\ \bibnamefont {Ewings}}, \bibinfo
  {author} {\bibfnamefont {J.~W.}\ \bibnamefont {Taylor}}, \bibinfo {author}
  {\bibfnamefont {A.~D.}\ \bibnamefont {Christianson}}, \ and\ \bibinfo
  {author} {\bibfnamefont {M.~A.}\ \bibnamefont {Green}},\ }\href@noop {}
  {\bibfield  {journal} {\bibinfo  {journal} {Phys. Rev. B}\ }\textbf {\bibinfo
  {volume} {90}} (\bibinfo {year} {2014})}\BibitemShut {NoStop}%
\bibitem [{\citenamefont {Oh}\ \emph {et~al.}(2013)\citenamefont {Oh},
  \citenamefont {Le}, \citenamefont {Jeong}, \citenamefont {Lee}, \citenamefont
  {Woo}, \citenamefont {Song}, \citenamefont {Perring}, \citenamefont {Buyers},
  \citenamefont {Cheong},\ and\ \citenamefont {Park}}]{Oh_PRL13}%
  \BibitemOpen
  \bibfield  {author} {\bibinfo {author} {\bibfnamefont {J.}~\bibnamefont
  {Oh}}, \bibinfo {author} {\bibfnamefont {M.~D.}\ \bibnamefont {Le}}, \bibinfo
  {author} {\bibfnamefont {J.}~\bibnamefont {Jeong}}, \bibinfo {author}
  {\bibfnamefont {J.-h.}\ \bibnamefont {Lee}}, \bibinfo {author} {\bibfnamefont
  {H.}~\bibnamefont {Woo}}, \bibinfo {author} {\bibfnamefont {W.-Y.}\
  \bibnamefont {Song}}, \bibinfo {author} {\bibfnamefont {T.~G.}\ \bibnamefont
  {Perring}}, \bibinfo {author} {\bibfnamefont {W.~J.~L.}\ \bibnamefont
  {Buyers}}, \bibinfo {author} {\bibfnamefont {S.-W.}\ \bibnamefont {Cheong}},
  \ and\ \bibinfo {author} {\bibfnamefont {J.-G.}\ \bibnamefont {Park}},\
  }\href@noop {} {\bibfield  {journal} {\bibinfo  {journal} {Phys. Rev. Lett.}\
  }\textbf {\bibinfo {volume} {111}},\ \bibinfo {pages} {257202} (\bibinfo
  {year} {2013})}\BibitemShut {NoStop}%
\bibitem [{\citenamefont {Schmidiger}\ \emph {et~al.}(2013)\citenamefont
  {Schmidiger}, \citenamefont {Bouillot}, \citenamefont {Guidi}, \citenamefont
  {Bewley}, \citenamefont {Kollath}, \citenamefont {Giamarchi},\ and\
  \citenamefont {Zheludev}}]{Schmidiger_PRL13}%
  \BibitemOpen
  \bibfield  {author} {\bibinfo {author} {\bibfnamefont {D.}~\bibnamefont
  {Schmidiger}}, \bibinfo {author} {\bibfnamefont {P.}~\bibnamefont
  {Bouillot}}, \bibinfo {author} {\bibfnamefont {T.}~\bibnamefont {Guidi}},
  \bibinfo {author} {\bibfnamefont {R.}~\bibnamefont {Bewley}}, \bibinfo
  {author} {\bibfnamefont {C.}~\bibnamefont {Kollath}}, \bibinfo {author}
  {\bibfnamefont {T.}~\bibnamefont {Giamarchi}}, \ and\ \bibinfo {author}
  {\bibfnamefont {A.}~\bibnamefont {Zheludev}},\ }\href@noop {} {\bibfield
  {journal} {\bibinfo  {journal} {Phys. Rev. Lett.}\ }\textbf {\bibinfo
  {volume} {111}},\ \bibinfo {pages} {107202} (\bibinfo {year}
  {2013})}\BibitemShut {NoStop}%
\bibitem [{\citenamefont {Lake}\ \emph {et~al.}(2013)\citenamefont {Lake},
  \citenamefont {Tennant}, \citenamefont {Caux}, \citenamefont {Barthel},
  \citenamefont {Schollwoeck}, \citenamefont {Nagler},\ and\ \citenamefont
  {Frost}}]{Lake-PRL-13}%
  \BibitemOpen
  \bibfield  {author} {\bibinfo {author} {\bibfnamefont {B.}~\bibnamefont
  {Lake}}, \bibinfo {author} {\bibfnamefont {D.~A.}\ \bibnamefont {Tennant}},
  \bibinfo {author} {\bibfnamefont {J.~S.}\ \bibnamefont {Caux}}, \bibinfo
  {author} {\bibfnamefont {T.}~\bibnamefont {Barthel}}, \bibinfo {author}
  {\bibfnamefont {U.}~\bibnamefont {Schollwoeck}}, \bibinfo {author}
  {\bibfnamefont {S.~E.}\ \bibnamefont {Nagler}}, \ and\ \bibinfo {author}
  {\bibfnamefont {C.~D.}\ \bibnamefont {Frost}},\ }\href@noop {} {\bibfield
  {journal} {\bibinfo  {journal} {Phys. Rev. Lett.}\ }\textbf {\bibinfo
  {volume} {111}} (\bibinfo {year} {2013})}\BibitemShut {NoStop}%
\bibitem [{\citenamefont {Johnstone}\ \emph {et~al.}(2012)\citenamefont
  {Johnstone}, \citenamefont {Perring}, \citenamefont {Sikora}, \citenamefont
  {Prabhakaran},\ and\ \citenamefont {Boothroyd}}]{Johnstone-PRL-12}%
  \BibitemOpen
  \bibfield  {author} {\bibinfo {author} {\bibfnamefont {G.~E.}\ \bibnamefont
  {Johnstone}}, \bibinfo {author} {\bibfnamefont {T.~G.}\ \bibnamefont
  {Perring}}, \bibinfo {author} {\bibfnamefont {O.}~\bibnamefont {Sikora}},
  \bibinfo {author} {\bibfnamefont {D.}~\bibnamefont {Prabhakaran}}, \ and\
  \bibinfo {author} {\bibfnamefont {A.~T.}\ \bibnamefont {Boothroyd}},\
  }\href@noop {} {\bibfield  {journal} {\bibinfo  {journal} {Phys. Rev. Lett.}\
  }\textbf {\bibinfo {volume} {109}} (\bibinfo {year} {2012})}\BibitemShut
  {NoStop}%
\bibitem [{\citenamefont {Lipscombe}\ \emph {et~al.}(2011)\citenamefont
  {Lipscombe}, \citenamefont {Chen}, \citenamefont {Fang}, \citenamefont
  {Perring}, \citenamefont {Abernathy}, \citenamefont {Christianson},
  \citenamefont {Egami}, \citenamefont {Wang}, \citenamefont {Hu},\ and\
  \citenamefont {Dai}}]{Lipscombe-PRL-11}%
  \BibitemOpen
  \bibfield  {author} {\bibinfo {author} {\bibfnamefont {O.~J.}\ \bibnamefont
  {Lipscombe}}, \bibinfo {author} {\bibfnamefont {G.~F.}\ \bibnamefont {Chen}},
  \bibinfo {author} {\bibfnamefont {C.}~\bibnamefont {Fang}}, \bibinfo {author}
  {\bibfnamefont {T.~G.}\ \bibnamefont {Perring}}, \bibinfo {author}
  {\bibfnamefont {D.~L.}\ \bibnamefont {Abernathy}}, \bibinfo {author}
  {\bibfnamefont {A.~D.}\ \bibnamefont {Christianson}}, \bibinfo {author}
  {\bibfnamefont {T.}~\bibnamefont {Egami}}, \bibinfo {author} {\bibfnamefont
  {N.}~\bibnamefont {Wang}}, \bibinfo {author} {\bibfnamefont {J.}~\bibnamefont
  {Hu}}, \ and\ \bibinfo {author} {\bibfnamefont {P.}~\bibnamefont {Dai}},\
  }\href@noop {} {\bibfield  {journal} {\bibinfo  {journal} {Phys. Rev. Lett.}\
  }\textbf {\bibinfo {volume} {106}} (\bibinfo {year} {2011})}\BibitemShut
  {NoStop}%
\bibitem [{\citenamefont {Headings}\ \emph {et~al.}(2010)\citenamefont
  {Headings}, \citenamefont {Hayden}, \citenamefont {Coldea},\ and\
  \citenamefont {Perring}}]{Headings_PRL10}%
  \BibitemOpen
  \bibfield  {author} {\bibinfo {author} {\bibfnamefont {N.~S.}\ \bibnamefont
  {Headings}}, \bibinfo {author} {\bibfnamefont {S.~M.}\ \bibnamefont
  {Hayden}}, \bibinfo {author} {\bibfnamefont {R.}~\bibnamefont {Coldea}}, \
  and\ \bibinfo {author} {\bibfnamefont {T.~G.}\ \bibnamefont {Perring}},\
  }\href@noop {} {\bibfield  {journal} {\bibinfo  {journal} {Phys. Rev. Lett.}\
  }\textbf {\bibinfo {volume} {105}},\ \bibinfo {pages} {247001} (\bibinfo
  {year} {2010})}\BibitemShut {NoStop}%
\bibitem [{\citenamefont {Doubble}\ \emph {et~al.}(2010)\citenamefont
  {Doubble}, \citenamefont {Hayden}, \citenamefont {Dai}, \citenamefont {Mook},
  \citenamefont {Thompson},\ and\ \citenamefont {Frost}}]{Doubble-PRL-10}%
  \BibitemOpen
  \bibfield  {author} {\bibinfo {author} {\bibfnamefont {R.}~\bibnamefont
  {Doubble}}, \bibinfo {author} {\bibfnamefont {S.~M.}\ \bibnamefont {Hayden}},
  \bibinfo {author} {\bibfnamefont {P.}~\bibnamefont {Dai}}, \bibinfo {author}
  {\bibfnamefont {H.~A.}\ \bibnamefont {Mook}}, \bibinfo {author}
  {\bibfnamefont {J.~R.}\ \bibnamefont {Thompson}}, \ and\ \bibinfo {author}
  {\bibfnamefont {C.~D.}\ \bibnamefont {Frost}},\ }\href@noop {} {\bibfield
  {journal} {\bibinfo  {journal} {Phys. Rev. Lett.}\ }\textbf {\bibinfo
  {volume} {105}} (\bibinfo {year} {2010})}\BibitemShut {NoStop}%
\bibitem [{\citenamefont {Lake}\ \emph {et~al.}(2010)\citenamefont {Lake},
  \citenamefont {Tsvelik}, \citenamefont {Notbohm}, \citenamefont {Tennant},
  \citenamefont {Perring}, \citenamefont {Reehuis}, \citenamefont {Sekar},
  \citenamefont {Krabbes},\ and\ \citenamefont {Buechner}}]{Lake-NatPhys-10}%
  \BibitemOpen
  \bibfield  {author} {\bibinfo {author} {\bibfnamefont {B.}~\bibnamefont
  {Lake}}, \bibinfo {author} {\bibfnamefont {A.~M.}\ \bibnamefont {Tsvelik}},
  \bibinfo {author} {\bibfnamefont {S.}~\bibnamefont {Notbohm}}, \bibinfo
  {author} {\bibfnamefont {D.~A.}\ \bibnamefont {Tennant}}, \bibinfo {author}
  {\bibfnamefont {T.~G.}\ \bibnamefont {Perring}}, \bibinfo {author}
  {\bibfnamefont {M.}~\bibnamefont {Reehuis}}, \bibinfo {author} {\bibfnamefont
  {C.}~\bibnamefont {Sekar}}, \bibinfo {author} {\bibfnamefont
  {G.}~\bibnamefont {Krabbes}}, \ and\ \bibinfo {author} {\bibfnamefont
  {B.}~\bibnamefont {Buechner}},\ }\href@noop {} {\bibfield  {journal}
  {\bibinfo  {journal} {Nat. Phys.}\ }\textbf {\bibinfo {volume} {6}},\
  \bibinfo {pages} {50} (\bibinfo {year} {2010})}\BibitemShut {NoStop}%
\bibitem [{\citenamefont {Walters}\ \emph {et~al.}(2009)\citenamefont
  {Walters}, \citenamefont {Perring}, \citenamefont {Caux}, \citenamefont
  {Savici}, \citenamefont {Gu}, \citenamefont {Lee}, \citenamefont {Ku},\ and\
  \citenamefont {Zaliznyak}}]{Walters-NPhys-09}%
  \BibitemOpen
  \bibfield  {author} {\bibinfo {author} {\bibfnamefont {A.~C.}\ \bibnamefont
  {Walters}}, \bibinfo {author} {\bibfnamefont {T.~G.}\ \bibnamefont
  {Perring}}, \bibinfo {author} {\bibfnamefont {J.-S.}\ \bibnamefont {Caux}},
  \bibinfo {author} {\bibfnamefont {A.~T.}\ \bibnamefont {Savici}}, \bibinfo
  {author} {\bibfnamefont {G.~D.}\ \bibnamefont {Gu}}, \bibinfo {author}
  {\bibfnamefont {C.-C.}\ \bibnamefont {Lee}}, \bibinfo {author} {\bibfnamefont
  {W.}~\bibnamefont {Ku}}, \ and\ \bibinfo {author} {\bibfnamefont {I.~A.}\
  \bibnamefont {Zaliznyak}},\ }\href@noop {} {\bibfield  {journal} {\bibinfo
  {journal} {Nat. Phys.}\ }\textbf {\bibinfo {volume} {5}},\ \bibinfo {pages}
  {867} (\bibinfo {year} {2009})}\BibitemShut {NoStop}%
\bibitem [{\citenamefont {Diallo}\ \emph {et~al.}(2009)\citenamefont {Diallo},
  \citenamefont {Antropov}, \citenamefont {Perring}, \citenamefont {Broholm},
  \citenamefont {Pulikkotil}, \citenamefont {Ni}, \citenamefont {Bud'ko},
  \citenamefont {Canfield}, \citenamefont {Kreyssig}, \citenamefont {Goldman},\
  and\ \citenamefont {McQueeney}}]{Diallo-PRL-09}%
  \BibitemOpen
  \bibfield  {author} {\bibinfo {author} {\bibfnamefont {S.~O.}\ \bibnamefont
  {Diallo}}, \bibinfo {author} {\bibfnamefont {V.~P.}\ \bibnamefont
  {Antropov}}, \bibinfo {author} {\bibfnamefont {T.~G.}\ \bibnamefont
  {Perring}}, \bibinfo {author} {\bibfnamefont {C.}~\bibnamefont {Broholm}},
  \bibinfo {author} {\bibfnamefont {J.~J.}\ \bibnamefont {Pulikkotil}},
  \bibinfo {author} {\bibfnamefont {N.}~\bibnamefont {Ni}}, \bibinfo {author}
  {\bibfnamefont {S.~L.}\ \bibnamefont {Bud'ko}}, \bibinfo {author}
  {\bibfnamefont {P.~C.}\ \bibnamefont {Canfield}}, \bibinfo {author}
  {\bibfnamefont {A.}~\bibnamefont {Kreyssig}}, \bibinfo {author}
  {\bibfnamefont {A.~I.}\ \bibnamefont {Goldman}}, \ and\ \bibinfo {author}
  {\bibfnamefont {R.~J.}\ \bibnamefont {McQueeney}},\ }\href@noop {} {\bibfield
   {journal} {\bibinfo  {journal} {Phys. Rev. Lett.}\ }\textbf {\bibinfo
  {volume} {102}} (\bibinfo {year} {2009})}\BibitemShut {NoStop}%
\bibitem [{\citenamefont {Lipscombe}\ \emph {et~al.}(2009)\citenamefont
  {Lipscombe}, \citenamefont {Vignolle}, \citenamefont {Perring}, \citenamefont
  {Frost},\ and\ \citenamefont {Hayden}}]{Lipscombe-PRL-09}%
  \BibitemOpen
  \bibfield  {author} {\bibinfo {author} {\bibfnamefont {O.~J.}\ \bibnamefont
  {Lipscombe}}, \bibinfo {author} {\bibfnamefont {B.}~\bibnamefont {Vignolle}},
  \bibinfo {author} {\bibfnamefont {T.~G.}\ \bibnamefont {Perring}}, \bibinfo
  {author} {\bibfnamefont {C.~D.}\ \bibnamefont {Frost}}, \ and\ \bibinfo
  {author} {\bibfnamefont {S.~M.}\ \bibnamefont {Hayden}},\ }\href@noop {}
  {\bibfield  {journal} {\bibinfo  {journal} {Phys. Rev. Lett.}\ }\textbf
  {\bibinfo {volume} {102}} (\bibinfo {year} {2009})}\BibitemShut {NoStop}%
\bibitem [{\citenamefont {Xu}\ \emph {et~al.}(2009)\citenamefont {Xu},
  \citenamefont {Gu}, \citenamefont {H\"{u}cker}, \citenamefont {Fauqu\'{e}},
  \citenamefont {Perring}, \citenamefont {Regnault},\ and\ \citenamefont
  {Tranquada}}]{Xu_NPhys09}%
  \BibitemOpen
  \bibfield  {author} {\bibinfo {author} {\bibfnamefont {G.}~\bibnamefont
  {Xu}}, \bibinfo {author} {\bibfnamefont {G.~D.}\ \bibnamefont {Gu}}, \bibinfo
  {author} {\bibfnamefont {M.}~\bibnamefont {H\"{u}cker}}, \bibinfo {author}
  {\bibfnamefont {B.}~\bibnamefont {Fauqu\'{e}}}, \bibinfo {author}
  {\bibfnamefont {T.~G.}\ \bibnamefont {Perring}}, \bibinfo {author}
  {\bibfnamefont {L.~P.}\ \bibnamefont {Regnault}}, \ and\ \bibinfo {author}
  {\bibfnamefont {J.~M.}\ \bibnamefont {Tranquada}},\ }\href@noop {} {\bibfield
   {journal} {\bibinfo  {journal} {Nat. Phys.}\ }\textbf {\bibinfo {volume}
  {5}},\ \bibinfo {pages} {642} (\bibinfo {year} {2009})}\BibitemShut {NoStop}%
\bibitem [{\citenamefont {Vignolle}\ \emph {et~al.}(2007)\citenamefont
  {Vignolle}, \citenamefont {Hayden}, \citenamefont {McMorrow}, \citenamefont
  {R{\o}nnow}, \citenamefont {Lake}, \citenamefont {Frost},\ and\ \citenamefont
  {Perring}}]{Vignolle_NPhys07}%
  \BibitemOpen
  \bibfield  {author} {\bibinfo {author} {\bibfnamefont {B.}~\bibnamefont
  {Vignolle}}, \bibinfo {author} {\bibfnamefont {S.~M.}\ \bibnamefont
  {Hayden}}, \bibinfo {author} {\bibfnamefont {D.~F.}\ \bibnamefont
  {McMorrow}}, \bibinfo {author} {\bibfnamefont {H.~M.}\ \bibnamefont
  {R{\o}nnow}}, \bibinfo {author} {\bibfnamefont {B.}~\bibnamefont {Lake}},
  \bibinfo {author} {\bibfnamefont {C.~D.}\ \bibnamefont {Frost}}, \ and\
  \bibinfo {author} {\bibfnamefont {T.~G.}\ \bibnamefont {Perring}},\
  }\href@noop {} {\bibfield  {journal} {\bibinfo  {journal} {Nat. Phys.}\
  }\textbf {\bibinfo {volume} {3}},\ \bibinfo {pages} {163 } (\bibinfo {year}
  {2007})}\BibitemShut {NoStop}%
\bibitem [{\citenamefont {Hayden}\ \emph {et~al.}(2004)\citenamefont {Hayden},
  \citenamefont {Mook}, \citenamefont {Dai}, \citenamefont {Perring},\ and\
  \citenamefont {Dogan}}]{Hayden-Nature}%
  \BibitemOpen
  \bibfield  {author} {\bibinfo {author} {\bibfnamefont {S.~M.}\ \bibnamefont
  {Hayden}}, \bibinfo {author} {\bibfnamefont {H.~A.}\ \bibnamefont {Mook}},
  \bibinfo {author} {\bibfnamefont {P.}~\bibnamefont {Dai}}, \bibinfo {author}
  {\bibfnamefont {T.~G.}\ \bibnamefont {Perring}}, \ and\ \bibinfo {author}
  {\bibfnamefont {F.}~\bibnamefont {Dogan}},\ }\href@noop {} {\bibfield
  {journal} {\bibinfo  {journal} {Nature}\ }\textbf {\bibinfo {volume} {429}},\
  \bibinfo {pages} {531} (\bibinfo {year} {2004})}\BibitemShut {NoStop}%
\bibitem [{\citenamefont {Tranquada}\ \emph {et~al.}(2004)\citenamefont
  {Tranquada}, \citenamefont {Woo}, \citenamefont {Perring}, \citenamefont
  {Goka}, \citenamefont {Gu}, \citenamefont {Xu}, \citenamefont {Fujita},\ and\
  \citenamefont {Yamada}}]{Tranquada-Nature}%
  \BibitemOpen
  \bibfield  {author} {\bibinfo {author} {\bibfnamefont {J.~M.}\ \bibnamefont
  {Tranquada}}, \bibinfo {author} {\bibfnamefont {H.}~\bibnamefont {Woo}},
  \bibinfo {author} {\bibfnamefont {T.~G.}\ \bibnamefont {Perring}}, \bibinfo
  {author} {\bibfnamefont {H.}~\bibnamefont {Goka}}, \bibinfo {author}
  {\bibfnamefont {G.~D.}\ \bibnamefont {Gu}}, \bibinfo {author} {\bibfnamefont
  {G.}~\bibnamefont {Xu}}, \bibinfo {author} {\bibfnamefont {M.}~\bibnamefont
  {Fujita}}, \ and\ \bibinfo {author} {\bibfnamefont {K.}~\bibnamefont
  {Yamada}},\ }\href@noop {} {\bibfield  {journal} {\bibinfo  {journal}
  {Nature}\ }\textbf {\bibinfo {volume} {429}},\ \bibinfo {pages} {534}
  (\bibinfo {year} {2004})}\BibitemShut {NoStop}%
\bibitem [{\citenamefont {Perring}\ \emph {et~al.}(1996)\citenamefont
  {Perring}, \citenamefont {Aeppli}, \citenamefont {Hayden}, \citenamefont
  {Carter}, \citenamefont {Remeika},\ and\ \citenamefont
  {Cheong}}]{Perring_manganite}%
  \BibitemOpen
  \bibfield  {author} {\bibinfo {author} {\bibfnamefont {T.~G.}\ \bibnamefont
  {Perring}}, \bibinfo {author} {\bibfnamefont {G.}~\bibnamefont {Aeppli}},
  \bibinfo {author} {\bibfnamefont {S.~M.}\ \bibnamefont {Hayden}}, \bibinfo
  {author} {\bibfnamefont {S.~A.}\ \bibnamefont {Carter}}, \bibinfo {author}
  {\bibfnamefont {J.~P.}\ \bibnamefont {Remeika}}, \ and\ \bibinfo {author}
  {\bibfnamefont {S.-W.}\ \bibnamefont {Cheong}},\ }\href@noop {} {\bibfield
  {journal} {\bibinfo  {journal} {Phys. Rev. Lett.}\ }\textbf {\bibinfo
  {volume} {77}},\ \bibinfo {pages} {711} (\bibinfo {year} {1996})}\BibitemShut
  {NoStop}%
\bibitem [{\citenamefont {Hayden}\ \emph {et~al.}(2000)\citenamefont {Hayden},
  \citenamefont {Doubble}, \citenamefont {Aeppli}, \citenamefont {Perring},\
  and\ \citenamefont {Fawcett}}]{Hayden_CrV}%
  \BibitemOpen
  \bibfield  {author} {\bibinfo {author} {\bibfnamefont {S.~M.}\ \bibnamefont
  {Hayden}}, \bibinfo {author} {\bibfnamefont {R.}~\bibnamefont {Doubble}},
  \bibinfo {author} {\bibfnamefont {G.}~\bibnamefont {Aeppli}}, \bibinfo
  {author} {\bibfnamefont {T.~G.}\ \bibnamefont {Perring}}, \ and\ \bibinfo
  {author} {\bibfnamefont {E.}~\bibnamefont {Fawcett}},\ }\href@noop {}
  {\bibfield  {journal} {\bibinfo  {journal} {Phys. Rev. Lett.}\ }\textbf
  {\bibinfo {volume} {84}},\ \bibinfo {pages} {999} (\bibinfo {year}
  {2000})}\BibitemShut {NoStop}%
\bibitem [{\citenamefont {Parker}\ \emph {et~al.}(2018)\citenamefont {Parker},
  \citenamefont {Ramirez-Cuesta},\ and\ \citenamefont
  {Daemen}}]{Parker-SpecActa-18}%
  \BibitemOpen
  \bibfield  {author} {\bibinfo {author} {\bibfnamefont {S.~F.}\ \bibnamefont
  {Parker}}, \bibinfo {author} {\bibfnamefont {A.~J.}\ \bibnamefont
  {Ramirez-Cuesta}}, \ and\ \bibinfo {author} {\bibfnamefont {L.}~\bibnamefont
  {Daemen}},\ }\href@noop {} {\bibfield  {journal} {\bibinfo  {journal}
  {Spectrochim. Acta A}\ }\textbf {\bibinfo {volume} {190}},\ \bibinfo {pages}
  {518} (\bibinfo {year} {2018})}\BibitemShut {NoStop}%
\bibitem [{\citenamefont {Brown}\ \emph {et~al.}(2017)\citenamefont {Brown},
  \citenamefont {Parker}, \citenamefont {Garcia}, \citenamefont {Mukhopadhyay},
  \citenamefont {Sakai},\ and\ \citenamefont {Stock}}]{Brown-PRB-17}%
  \BibitemOpen
  \bibfield  {author} {\bibinfo {author} {\bibfnamefont {K.~L.}\ \bibnamefont
  {Brown}}, \bibinfo {author} {\bibfnamefont {S.~F.}\ \bibnamefont {Parker}},
  \bibinfo {author} {\bibfnamefont {I.~R.}\ \bibnamefont {Garcia}}, \bibinfo
  {author} {\bibfnamefont {S.}~\bibnamefont {Mukhopadhyay}}, \bibinfo {author}
  {\bibfnamefont {V.~G.}\ \bibnamefont {Sakai}}, \ and\ \bibinfo {author}
  {\bibfnamefont {C.}~\bibnamefont {Stock}},\ }\href@noop {} {\bibfield
  {journal} {\bibinfo  {journal} {Phys. Rev. B}\ }\textbf {\bibinfo {volume}
  {96}} (\bibinfo {year} {2017})}\BibitemShut {NoStop}%
\bibitem [{\citenamefont {O'Malley}\ \emph {et~al.}(2017)\citenamefont
  {O'Malley}, \citenamefont {Parker},\ and\ \citenamefont
  {Catlow}}]{OMalley-ChemCom-17}%
  \BibitemOpen
  \bibfield  {author} {\bibinfo {author} {\bibfnamefont {A.~J.}\ \bibnamefont
  {O'Malley}}, \bibinfo {author} {\bibfnamefont {S.~F.}\ \bibnamefont
  {Parker}}, \ and\ \bibinfo {author} {\bibfnamefont {C.~R.~A.}\ \bibnamefont
  {Catlow}},\ }\href@noop {} {\bibfield  {journal} {\bibinfo  {journal} {Chem.
  Comm.}\ }\textbf {\bibinfo {volume} {53}},\ \bibinfo {pages} {12164}
  (\bibinfo {year} {2017})}\BibitemShut {NoStop}%
\bibitem [{\citenamefont {Albers}\ \emph {et~al.}(2016)\citenamefont {Albers},
  \citenamefont {Weber}, \citenamefont {Moebus}, \citenamefont {Wieland},\ and\
  \citenamefont {Parker}}]{Albers-carbon-16}%
  \BibitemOpen
  \bibfield  {author} {\bibinfo {author} {\bibfnamefont {P.~W.}\ \bibnamefont
  {Albers}}, \bibinfo {author} {\bibfnamefont {W.}~\bibnamefont {Weber}},
  \bibinfo {author} {\bibfnamefont {K.}~\bibnamefont {Moebus}}, \bibinfo
  {author} {\bibfnamefont {S.~D.}\ \bibnamefont {Wieland}}, \ and\ \bibinfo
  {author} {\bibfnamefont {S.~F.}\ \bibnamefont {Parker}},\ }\href@noop {}
  {\bibfield  {journal} {\bibinfo  {journal} {Carbon}\ }\textbf {\bibinfo
  {volume} {109}},\ \bibinfo {pages} {239} (\bibinfo {year}
  {2016})}\BibitemShut {NoStop}%
\bibitem [{\citenamefont {Cavallari}\ \emph {et~al.}(2016)\citenamefont
  {Cavallari}, \citenamefont {Pontiroli}, \citenamefont {Jimenez-Ruiz},
  \citenamefont {Johnson}, \citenamefont {Aramini}, \citenamefont {Gaboardi},
  \citenamefont {Parker}, \citenamefont {Ricco},\ and\ \citenamefont
  {Rols}}]{Cavallari-PhysChemChemPhys-16}%
  \BibitemOpen
  \bibfield  {author} {\bibinfo {author} {\bibfnamefont {C.}~\bibnamefont
  {Cavallari}}, \bibinfo {author} {\bibfnamefont {D.}~\bibnamefont
  {Pontiroli}}, \bibinfo {author} {\bibfnamefont {M.}~\bibnamefont
  {Jimenez-Ruiz}}, \bibinfo {author} {\bibfnamefont {M.}~\bibnamefont
  {Johnson}}, \bibinfo {author} {\bibfnamefont {M.}~\bibnamefont {Aramini}},
  \bibinfo {author} {\bibfnamefont {M.}~\bibnamefont {Gaboardi}}, \bibinfo
  {author} {\bibfnamefont {S.~F.}\ \bibnamefont {Parker}}, \bibinfo {author}
  {\bibfnamefont {M.}~\bibnamefont {Ricco}}, \ and\ \bibinfo {author}
  {\bibfnamefont {S.}~\bibnamefont {Rols}},\ }\href@noop {} {\bibfield
  {journal} {\bibinfo  {journal} {Phys. Chem. Chem. Phys.}\ }\textbf {\bibinfo
  {volume} {18}},\ \bibinfo {pages} {24820} (\bibinfo {year}
  {2016})}\BibitemShut {NoStop}%
\bibitem [{\citenamefont {Parker}\ and\ \citenamefont
  {Collier}(2016)}]{Parker-JM-Rev-16}%
  \BibitemOpen
  \bibfield  {author} {\bibinfo {author} {\bibfnamefont {S.~F.}\ \bibnamefont
  {Parker}}\ and\ \bibinfo {author} {\bibfnamefont {P.}~\bibnamefont
  {Collier}},\ }\href@noop {} {\bibfield  {journal} {\bibinfo  {journal}
  {Johnson Matthey Technology Review}\ }\textbf {\bibinfo {volume} {60}},\
  \bibinfo {pages} {132} (\bibinfo {year} {2016})}\BibitemShut {NoStop}%
\bibitem [{\citenamefont {Marques}\ \emph {et~al.}(2016)\citenamefont
  {Marques}, \citenamefont {Goncalves}, \citenamefont {Amarante}, \citenamefont
  {Makhoul}, \citenamefont {Parker},\ and\ \citenamefont {Batista~de
  Carvalho}}]{Marques-bone-16}%
  \BibitemOpen
  \bibfield  {author} {\bibinfo {author} {\bibfnamefont {M.~P.~M.}\
  \bibnamefont {Marques}}, \bibinfo {author} {\bibfnamefont {D.}~\bibnamefont
  {Goncalves}}, \bibinfo {author} {\bibfnamefont {A.~I.~C.}\ \bibnamefont
  {Amarante}}, \bibinfo {author} {\bibfnamefont {C.~I.}\ \bibnamefont
  {Makhoul}}, \bibinfo {author} {\bibfnamefont {S.~F.}\ \bibnamefont {Parker}},
  \ and\ \bibinfo {author} {\bibfnamefont {L.~A.~E.}\ \bibnamefont {Batista~de
  Carvalho}},\ }\href@noop {} {\bibfield  {journal} {\bibinfo  {journal} {RSC
  Advances}\ }\textbf {\bibinfo {volume} {6}},\ \bibinfo {pages} {68638}
  (\bibinfo {year} {2016})}\BibitemShut {NoStop}%
\bibitem [{\citenamefont {Warringham}\ \emph {et~al.}(2015)\citenamefont
  {Warringham}, \citenamefont {McFarlane}, \citenamefont {MacLaren},
  \citenamefont {Webb}, \citenamefont {Tooze}, \citenamefont {Taylor},
  \citenamefont {Ewings}, \citenamefont {Parker},\ and\ \citenamefont
  {Lennon}}]{Warringham-JChemPhys}%
  \BibitemOpen
  \bibfield  {author} {\bibinfo {author} {\bibfnamefont {R.}~\bibnamefont
  {Warringham}}, \bibinfo {author} {\bibfnamefont {A.~R.}\ \bibnamefont
  {McFarlane}}, \bibinfo {author} {\bibfnamefont {D.~A.}\ \bibnamefont
  {MacLaren}}, \bibinfo {author} {\bibfnamefont {P.~B.}\ \bibnamefont {Webb}},
  \bibinfo {author} {\bibfnamefont {R.~P.}\ \bibnamefont {Tooze}}, \bibinfo
  {author} {\bibfnamefont {J.}~\bibnamefont {Taylor}}, \bibinfo {author}
  {\bibfnamefont {R.~A.}\ \bibnamefont {Ewings}}, \bibinfo {author}
  {\bibfnamefont {S.~F.}\ \bibnamefont {Parker}}, \ and\ \bibinfo {author}
  {\bibfnamefont {D.}~\bibnamefont {Lennon}},\ }\href@noop {} {\bibfield
  {journal} {\bibinfo  {journal} {J. Chem. Phys.}\ }\textbf {\bibinfo {volume}
  {143}} (\bibinfo {year} {2015})}\BibitemShut {NoStop}%
\bibitem [{\citenamefont {Albers}\ \emph {et~al.}(2015)\citenamefont {Albers},
  \citenamefont {Moebus}, \citenamefont {Wieland},\ and\ \citenamefont
  {Parker}}]{Albers-Pearlmans-15}%
  \BibitemOpen
  \bibfield  {author} {\bibinfo {author} {\bibfnamefont {P.~W.}\ \bibnamefont
  {Albers}}, \bibinfo {author} {\bibfnamefont {K.}~\bibnamefont {Moebus}},
  \bibinfo {author} {\bibfnamefont {S.~D.}\ \bibnamefont {Wieland}}, \ and\
  \bibinfo {author} {\bibfnamefont {S.~F.}\ \bibnamefont {Parker}},\
  }\href@noop {} {\bibfield  {journal} {\bibinfo  {journal} {Phys. Chem. Chem.
  Phys.}\ }\textbf {\bibinfo {volume} {17}},\ \bibinfo {pages} {5274} (\bibinfo
  {year} {2015})}\BibitemShut {NoStop}%
\bibitem [{\citenamefont {Parker}\ \emph {et~al.}(2014)\citenamefont {Parker},
  \citenamefont {Ramirez-Cuesta}, \citenamefont {Albers},\ and\ \citenamefont
  {Lennon}}]{Parker-review}%
  \BibitemOpen
  \bibfield  {author} {\bibinfo {author} {\bibfnamefont {S.~F.}\ \bibnamefont
  {Parker}}, \bibinfo {author} {\bibfnamefont {A.~J.}\ \bibnamefont
  {Ramirez-Cuesta}}, \bibinfo {author} {\bibfnamefont {P.~W.}\ \bibnamefont
  {Albers}}, \ and\ \bibinfo {author} {\bibfnamefont {D.}~\bibnamefont
  {Lennon}},\ }in\ \href@noop {} {\emph {\bibinfo {booktitle} {Dynamics of
  molecules and materials-II}}},\ \bibinfo {series} {Journal of Physics
  Conference Series}, Vol.\ \bibinfo {volume} {554},\ \bibinfo {editor} {edited
  by\ \bibinfo {editor} {\bibfnamefont {M.}~\bibnamefont {JimenezRuiz}}\ and\
  \bibinfo {editor} {\bibfnamefont {S.}~\bibnamefont {Parker}}}\ (\bibinfo
  {year} {2014})\BibitemShut {NoStop}%
\bibitem [{\citenamefont {Bewley}\ \emph {et~al.}(2006)\citenamefont {Bewley},
  \citenamefont {Eccleston}, \citenamefont {McEwen}, \citenamefont {Hayden},
  \citenamefont {Dove}, \citenamefont {Bennington}, \citenamefont {Treadgold},\
  and\ \citenamefont {Coleman}}]{Bewley20061029}%
  \BibitemOpen
  \bibfield  {author} {\bibinfo {author} {\bibfnamefont {R.~I.}\ \bibnamefont
  {Bewley}}, \bibinfo {author} {\bibfnamefont {R.~S.}\ \bibnamefont
  {Eccleston}}, \bibinfo {author} {\bibfnamefont {K.~A.}\ \bibnamefont
  {McEwen}}, \bibinfo {author} {\bibfnamefont {S.~M.}\ \bibnamefont {Hayden}},
  \bibinfo {author} {\bibfnamefont {M.~T.}\ \bibnamefont {Dove}}, \bibinfo
  {author} {\bibfnamefont {S.~M.}\ \bibnamefont {Bennington}}, \bibinfo
  {author} {\bibfnamefont {J.~R.}\ \bibnamefont {Treadgold}}, \ and\ \bibinfo
  {author} {\bibfnamefont {R.~L.~S.}\ \bibnamefont {Coleman}},\ }\href@noop {}
  {\bibfield  {journal} {\bibinfo  {journal} {Physica B}\ }\textbf {\bibinfo
  {volume} {385}},\ \bibinfo {pages} {1029 } (\bibinfo {year}
  {2006})}\BibitemShut {NoStop}%
\bibitem [{\citenamefont {Bewley}\ \emph {et~al.}(2011)\citenamefont {Bewley},
  \citenamefont {Taylor},\ and\ \citenamefont {Bennington.}}]{Bewley-LET}%
  \BibitemOpen
  \bibfield  {author} {\bibinfo {author} {\bibfnamefont {R.}~\bibnamefont
  {Bewley}}, \bibinfo {author} {\bibfnamefont {J.}~\bibnamefont {Taylor}}, \
  and\ \bibinfo {author} {\bibfnamefont {S.}~\bibnamefont {Bennington.}},\
  }\href@noop {} {\bibfield  {journal} {\bibinfo  {journal} {Nuc. Inst. Meth.
  A}\ }\textbf {\bibinfo {volume} {637}},\ \bibinfo {pages} {128 } (\bibinfo
  {year} {2011})}\BibitemShut {NoStop}%
\bibitem [{\citenamefont {Abernathy}\ \emph {et~al.}(2012)\citenamefont
  {Abernathy}, \citenamefont {Stone}, \citenamefont {Loguillo}, \citenamefont
  {Lucas}, \citenamefont {Delaire}, \citenamefont {Tang}, \citenamefont {Lin},\
  and\ \citenamefont {Fultz}}]{Abernathy-2012}%
  \BibitemOpen
  \bibfield  {author} {\bibinfo {author} {\bibfnamefont {D.~L.}\ \bibnamefont
  {Abernathy}}, \bibinfo {author} {\bibfnamefont {M.~B.}\ \bibnamefont
  {Stone}}, \bibinfo {author} {\bibfnamefont {M.~J.}\ \bibnamefont {Loguillo}},
  \bibinfo {author} {\bibfnamefont {M.~S.}\ \bibnamefont {Lucas}}, \bibinfo
  {author} {\bibfnamefont {O.}~\bibnamefont {Delaire}}, \bibinfo {author}
  {\bibfnamefont {X.}~\bibnamefont {Tang}}, \bibinfo {author} {\bibfnamefont
  {J.~Y.~Y.}\ \bibnamefont {Lin}}, \ and\ \bibinfo {author} {\bibfnamefont
  {B.}~\bibnamefont {Fultz}},\ }\href@noop {} {\bibfield  {journal} {\bibinfo
  {journal} {Rev. Sci. Inst.}\ }\textbf {\bibinfo {volume} {83}},\ \bibinfo
  {pages} {015114} (\bibinfo {year} {2012})}\BibitemShut {NoStop}%
\bibitem [{\citenamefont {Granroth}\ \emph {et~al.}(2010)\citenamefont
  {Granroth}, \citenamefont {Kolesnikov}, \citenamefont {Sherline},
  \citenamefont {Clancy}, \citenamefont {Ross}, \citenamefont {Ruff},
  \citenamefont {Gaulin},\ and\ \citenamefont {Nagler}}]{Granroth-SEQ}%
  \BibitemOpen
  \bibfield  {author} {\bibinfo {author} {\bibfnamefont {G.~E.}\ \bibnamefont
  {Granroth}}, \bibinfo {author} {\bibfnamefont {A.~I.}\ \bibnamefont
  {Kolesnikov}}, \bibinfo {author} {\bibfnamefont {T.~E.}\ \bibnamefont
  {Sherline}}, \bibinfo {author} {\bibfnamefont {J.~P.}\ \bibnamefont
  {Clancy}}, \bibinfo {author} {\bibfnamefont {K.~A.}\ \bibnamefont {Ross}},
  \bibinfo {author} {\bibfnamefont {J.~P.~C.}\ \bibnamefont {Ruff}}, \bibinfo
  {author} {\bibfnamefont {B.~D.}\ \bibnamefont {Gaulin}}, \ and\ \bibinfo
  {author} {\bibfnamefont {S.~E.}\ \bibnamefont {Nagler}},\ }\href@noop {}
  {\bibfield  {journal} {\bibinfo  {journal} {J. Phys. Conf. Ser.}\ }\textbf
  {\bibinfo {volume} {251}},\ \bibinfo {pages} {012058} (\bibinfo {year}
  {2010})}\BibitemShut {NoStop}%
\bibitem [{\citenamefont {Ehlers}\ \emph {et~al.}(2011)\citenamefont {Ehlers},
  \citenamefont {Podlesnyak}, \citenamefont {Niedziela}, \citenamefont
  {Iverson},\ and\ \citenamefont {Sokol}}]{Ehlers-CNCS}%
  \BibitemOpen
  \bibfield  {author} {\bibinfo {author} {\bibfnamefont {G.}~\bibnamefont
  {Ehlers}}, \bibinfo {author} {\bibfnamefont {A.~A.}\ \bibnamefont
  {Podlesnyak}}, \bibinfo {author} {\bibfnamefont {J.~L.}\ \bibnamefont
  {Niedziela}}, \bibinfo {author} {\bibfnamefont {E.~B.}\ \bibnamefont
  {Iverson}}, \ and\ \bibinfo {author} {\bibfnamefont {P.}~\bibnamefont
  {Sokol}},\ }\href@noop {} {\bibfield  {journal} {\bibinfo  {journal} {Rev.
  Sci. Instrum.}\ }\textbf {\bibinfo {volume} {82}},\ \bibinfo {pages} {085108}
  (\bibinfo {year} {2011})}\BibitemShut {NoStop}%
\bibitem [{\citenamefont {Kajimoto}\ \emph {et~al.}(2007)\citenamefont
  {Kajimoto}, \citenamefont {Yokoo}, \citenamefont {Nakajima}, \citenamefont
  {Nakamura}, \citenamefont {Soyama}, \citenamefont {Ino}, \citenamefont
  {Shamoto}, \citenamefont {Fujita}, \citenamefont {Ohoyama}, \citenamefont
  {Hiraka}, \citenamefont {Yamada},\ and\ \citenamefont {Arai}}]{4seasons}%
  \BibitemOpen
  \bibfield  {author} {\bibinfo {author} {\bibfnamefont {R.}~\bibnamefont
  {Kajimoto}}, \bibinfo {author} {\bibfnamefont {T.}~\bibnamefont {Yokoo}},
  \bibinfo {author} {\bibfnamefont {K.}~\bibnamefont {Nakajima}}, \bibinfo
  {author} {\bibfnamefont {M.}~\bibnamefont {Nakamura}}, \bibinfo {author}
  {\bibfnamefont {K.}~\bibnamefont {Soyama}}, \bibinfo {author} {\bibfnamefont
  {T.}~\bibnamefont {Ino}}, \bibinfo {author} {\bibfnamefont {S.}~\bibnamefont
  {Shamoto}}, \bibinfo {author} {\bibfnamefont {M.}~\bibnamefont {Fujita}},
  \bibinfo {author} {\bibfnamefont {K.}~\bibnamefont {Ohoyama}}, \bibinfo
  {author} {\bibfnamefont {H.}~\bibnamefont {Hiraka}}, \bibinfo {author}
  {\bibfnamefont {K.}~\bibnamefont {Yamada}}, \ and\ \bibinfo {author}
  {\bibfnamefont {M.}~\bibnamefont {Arai}},\ }\href@noop {} {\bibfield
  {journal} {\bibinfo  {journal} {Journal of Neutron Research}\ }\textbf
  {\bibinfo {volume} {15}},\ \bibinfo {pages} {5} (\bibinfo {year}
  {2007})}\BibitemShut {NoStop}%
\bibitem [{\citenamefont {Itoh}\ \emph {et~al.}(2011)\citenamefont {Itoh},
  \citenamefont {Yokoo}, \citenamefont {Satoh}, \citenamefont {ichiro Yano},
  \citenamefont {Kawana}, \citenamefont {Suzuki},\ and\ \citenamefont
  {Sato}}]{Itoh-HRC}%
  \BibitemOpen
  \bibfield  {author} {\bibinfo {author} {\bibfnamefont {S.}~\bibnamefont
  {Itoh}}, \bibinfo {author} {\bibfnamefont {T.}~\bibnamefont {Yokoo}},
  \bibinfo {author} {\bibfnamefont {S.}~\bibnamefont {Satoh}}, \bibinfo
  {author} {\bibfnamefont {S.}~\bibnamefont {ichiro Yano}}, \bibinfo {author}
  {\bibfnamefont {D.}~\bibnamefont {Kawana}}, \bibinfo {author} {\bibfnamefont
  {J.}~\bibnamefont {Suzuki}}, \ and\ \bibinfo {author} {\bibfnamefont {T.~J.}\
  \bibnamefont {Sato}},\ }\href@noop {} {\bibfield  {journal} {\bibinfo
  {journal} {Nuc. Inst. Meth. A}\ }\textbf {\bibinfo {volume} {631}},\ \bibinfo
  {pages} {90 } (\bibinfo {year} {2011})}\BibitemShut {NoStop}%
\bibitem [{\citenamefont {Nakajima}\ \emph {et~al.}(2011)\citenamefont
  {Nakajima}, \citenamefont {Ohira-Kawamura}, \citenamefont {Kikuchi},
  \citenamefont {Nakamura}, \citenamefont {Kajimoto}, \citenamefont {Inamura},
  \citenamefont {Takahashi}, \citenamefont {Aizawa}, \citenamefont {Suzuya},
  \citenamefont {Shibata}, \citenamefont {Nakatani}, \citenamefont {Soyama},
  \citenamefont {Maruyama}, \citenamefont {Tanaka}, \citenamefont {Kambara},
  \citenamefont {Iwahashi}, \citenamefont {Itoh}, \citenamefont {Osakabe},
  \citenamefont {Wakimoto}, \citenamefont {Kakurai}, \citenamefont {Maekawa},
  \citenamefont {Harada}, \citenamefont {Oikawa}, \citenamefont {E.~Lechner},
  \citenamefont {Mezei},\ and\ \citenamefont {Arai}}]{Nakajima-AMATERAS}%
  \BibitemOpen
  \bibfield  {author} {\bibinfo {author} {\bibfnamefont {K.}~\bibnamefont
  {Nakajima}}, \bibinfo {author} {\bibfnamefont {S.}~\bibnamefont
  {Ohira-Kawamura}}, \bibinfo {author} {\bibfnamefont {T.}~\bibnamefont
  {Kikuchi}}, \bibinfo {author} {\bibfnamefont {M.}~\bibnamefont {Nakamura}},
  \bibinfo {author} {\bibfnamefont {R.}~\bibnamefont {Kajimoto}}, \bibinfo
  {author} {\bibfnamefont {Y.}~\bibnamefont {Inamura}}, \bibinfo {author}
  {\bibfnamefont {N.}~\bibnamefont {Takahashi}}, \bibinfo {author}
  {\bibfnamefont {K.}~\bibnamefont {Aizawa}}, \bibinfo {author} {\bibfnamefont
  {K.}~\bibnamefont {Suzuya}}, \bibinfo {author} {\bibfnamefont
  {K.}~\bibnamefont {Shibata}}, \bibinfo {author} {\bibfnamefont
  {T.}~\bibnamefont {Nakatani}}, \bibinfo {author} {\bibfnamefont
  {K.}~\bibnamefont {Soyama}}, \bibinfo {author} {\bibfnamefont
  {R.}~\bibnamefont {Maruyama}}, \bibinfo {author} {\bibfnamefont
  {H.}~\bibnamefont {Tanaka}}, \bibinfo {author} {\bibfnamefont
  {W.}~\bibnamefont {Kambara}}, \bibinfo {author} {\bibfnamefont
  {T.}~\bibnamefont {Iwahashi}}, \bibinfo {author} {\bibfnamefont
  {Y.}~\bibnamefont {Itoh}}, \bibinfo {author} {\bibfnamefont {T.}~\bibnamefont
  {Osakabe}}, \bibinfo {author} {\bibfnamefont {S.}~\bibnamefont {Wakimoto}},
  \bibinfo {author} {\bibfnamefont {K.}~\bibnamefont {Kakurai}}, \bibinfo
  {author} {\bibfnamefont {F.}~\bibnamefont {Maekawa}}, \bibinfo {author}
  {\bibfnamefont {M.}~\bibnamefont {Harada}}, \bibinfo {author} {\bibfnamefont
  {K.}~\bibnamefont {Oikawa}}, \bibinfo {author} {\bibfnamefont
  {R.}~\bibnamefont {E.~Lechner}}, \bibinfo {author} {\bibfnamefont
  {F.}~\bibnamefont {Mezei}}, \ and\ \bibinfo {author} {\bibfnamefont
  {M.}~\bibnamefont {Arai}},\ }\href@noop {} {\bibfield  {journal} {\bibinfo
  {journal} {J. Phys. Soc. Japan}\ }\textbf {\bibinfo {volume} {80}},\ \bibinfo
  {pages} {SB028} (\bibinfo {year} {2011})}\BibitemShut {NoStop}%
\bibitem [{\citenamefont {Ollivier}\ and\ \citenamefont
  {Mutka}(2011)}]{Ollivier-IN5}%
  \BibitemOpen
  \bibfield  {author} {\bibinfo {author} {\bibfnamefont {J.}~\bibnamefont
  {Ollivier}}\ and\ \bibinfo {author} {\bibfnamefont {H.}~\bibnamefont
  {Mutka}},\ }\href@noop {} {\bibfield  {journal} {\bibinfo  {journal} {J.
  Phys. Soc. Jpn}\ }\textbf {\bibinfo {volume} {80}},\ \bibinfo {pages} {SB003}
  (\bibinfo {year} {2011})}\BibitemShut {NoStop}%
\bibitem [{\citenamefont {Russina}\ \emph {et~al.}(2017)\citenamefont
  {Russina}, \citenamefont {Guenther}, \citenamefont {Grzimek}, \citenamefont
  {Gainov}, \citenamefont {Schlegel}, \citenamefont {Drescher}, \citenamefont
  {Kaulich}, \citenamefont {Graf}, \citenamefont {Urban}, \citenamefont
  {Daske}, \citenamefont {Grotjahn}, \citenamefont {Hellhammer}, \citenamefont
  {Buchert}, \citenamefont {Kutz}, \citenamefont {Rossa}, \citenamefont
  {Sauer}, \citenamefont {Fromme}, \citenamefont {Wallacher}, \citenamefont
  {Kiefer}, \citenamefont {Klemke}, \citenamefont {Grimm}, \citenamefont
  {Gerischer}, \citenamefont {Tsapatsaris},\ and\ \citenamefont
  {Rolfs}}]{RUSSINA2017}%
  \BibitemOpen
  \bibfield  {author} {\bibinfo {author} {\bibfnamefont {M.}~\bibnamefont
  {Russina}}, \bibinfo {author} {\bibfnamefont {G.}~\bibnamefont {Guenther}},
  \bibinfo {author} {\bibfnamefont {V.}~\bibnamefont {Grzimek}}, \bibinfo
  {author} {\bibfnamefont {R.}~\bibnamefont {Gainov}}, \bibinfo {author}
  {\bibfnamefont {M.-C.}\ \bibnamefont {Schlegel}}, \bibinfo {author}
  {\bibfnamefont {L.}~\bibnamefont {Drescher}}, \bibinfo {author}
  {\bibfnamefont {T.}~\bibnamefont {Kaulich}}, \bibinfo {author} {\bibfnamefont
  {W.}~\bibnamefont {Graf}}, \bibinfo {author} {\bibfnamefont {B.}~\bibnamefont
  {Urban}}, \bibinfo {author} {\bibfnamefont {A.}~\bibnamefont {Daske}},
  \bibinfo {author} {\bibfnamefont {K.}~\bibnamefont {Grotjahn}}, \bibinfo
  {author} {\bibfnamefont {R.}~\bibnamefont {Hellhammer}}, \bibinfo {author}
  {\bibfnamefont {G.}~\bibnamefont {Buchert}}, \bibinfo {author} {\bibfnamefont
  {H.}~\bibnamefont {Kutz}}, \bibinfo {author} {\bibfnamefont {L.}~\bibnamefont
  {Rossa}}, \bibinfo {author} {\bibfnamefont {O.-P.}\ \bibnamefont {Sauer}},
  \bibinfo {author} {\bibfnamefont {M.}~\bibnamefont {Fromme}}, \bibinfo
  {author} {\bibfnamefont {D.}~\bibnamefont {Wallacher}}, \bibinfo {author}
  {\bibfnamefont {K.}~\bibnamefont {Kiefer}}, \bibinfo {author} {\bibfnamefont
  {B.}~\bibnamefont {Klemke}}, \bibinfo {author} {\bibfnamefont
  {N.}~\bibnamefont {Grimm}}, \bibinfo {author} {\bibfnamefont
  {S.}~\bibnamefont {Gerischer}}, \bibinfo {author} {\bibfnamefont
  {N.}~\bibnamefont {Tsapatsaris}}, \ and\ \bibinfo {author} {\bibfnamefont
  {K.}~\bibnamefont {Rolfs}},\ }\href@noop {} {\bibfield  {journal} {\bibinfo
  {journal} {Physica B}\ } (\bibinfo {year} {2017})}\BibitemShut {NoStop}%
\bibitem [{\citenamefont {\v{S}koro}\ \emph {et~al.}(2018)\citenamefont
  {\v{S}koro}, \citenamefont {Bewley}, \citenamefont {Lilley}, \citenamefont
  {Ewings}, \citenamefont {Romanelli}, \citenamefont {Gutmann}, \citenamefont
  {Smith}, \citenamefont {Rudi\'{c}},\ and\ \citenamefont
  {Ansell}}]{Skoro-moderators}%
  \BibitemOpen
  \bibfield  {author} {\bibinfo {author} {\bibfnamefont {G.}~\bibnamefont
  {\v{S}koro}}, \bibinfo {author} {\bibfnamefont {R.}~\bibnamefont {Bewley}},
  \bibinfo {author} {\bibfnamefont {S.}~\bibnamefont {Lilley}}, \bibinfo
  {author} {\bibfnamefont {R.}~\bibnamefont {Ewings}}, \bibinfo {author}
  {\bibfnamefont {G.}~\bibnamefont {Romanelli}}, \bibinfo {author}
  {\bibfnamefont {M.}~\bibnamefont {Gutmann}}, \bibinfo {author} {\bibfnamefont
  {R.}~\bibnamefont {Smith}}, \bibinfo {author} {\bibfnamefont
  {S.}~\bibnamefont {Rudi\'{c}}}, \ and\ \bibinfo {author} {\bibfnamefont
  {S.}~\bibnamefont {Ansell}},\ }\href@noop {} {\bibfield  {journal} {\bibinfo
  {journal} {J. Phys. Conf. Ser.}\ }\textbf {\bibinfo {volume} {1021}},\
  \bibinfo {pages} {012039} (\bibinfo {year} {2018})}\BibitemShut {NoStop}%
\bibitem [{\citenamefont {Fermi}\ \emph {et~al.}(1947)\citenamefont {Fermi},
  \citenamefont {Marshall},\ and\ \citenamefont {Marshall}}]{Fermi-orig}%
  \BibitemOpen
  \bibfield  {author} {\bibinfo {author} {\bibfnamefont {E.}~\bibnamefont
  {Fermi}}, \bibinfo {author} {\bibfnamefont {J.}~\bibnamefont {Marshall}}, \
  and\ \bibinfo {author} {\bibfnamefont {L.}~\bibnamefont {Marshall}},\
  }\href@noop {} {\bibfield  {journal} {\bibinfo  {journal} {Phys. Rev.}\
  }\textbf {\bibinfo {volume} {72}},\ \bibinfo {pages} {193} (\bibinfo {year}
  {1947})}\BibitemShut {NoStop}%
\bibitem [{VAT()}]{VAT-ref}%
  \BibitemOpen
  \href@noop {} {}\bibinfo {howpublished}
  {\url{https://www.lesker.com/newweb/valves/vat/}},\ \bibinfo {note}
  {accessed: 06/07/2018}\BibitemShut {NoStop}%
\bibitem [{\citenamefont {Voitovetskii}\ \emph {et~al.}(1959)\citenamefont
  {Voitovetskii}, \citenamefont {Tolmacheva},\ and\ \citenamefont
  {Arsaev}}]{Monitor-1}%
  \BibitemOpen
  \bibfield  {author} {\bibinfo {author} {\bibfnamefont {V.~K.}\ \bibnamefont
  {Voitovetskii}}, \bibinfo {author} {\bibfnamefont {N.~S.}\ \bibnamefont
  {Tolmacheva}}, \ and\ \bibinfo {author} {\bibfnamefont {M.~I.}\ \bibnamefont
  {Arsaev}},\ }\href@noop {} {\bibfield  {journal} {\bibinfo  {journal} {Atomn.
  Energ.}\ }\textbf {\bibinfo {volume} {6}},\ \bibinfo {pages} {321} (\bibinfo
  {year} {1959})}\BibitemShut {NoStop}%
\bibitem [{\citenamefont {Spoward}(1969)}]{Monitor-2}%
  \BibitemOpen
  \bibfield  {author} {\bibinfo {author} {\bibfnamefont {A.~R.}\ \bibnamefont
  {Spoward}},\ }\href@noop {} {\bibfield  {journal} {\bibinfo  {journal} {Nucl
  Instr. and Meth.}\ }\textbf {\bibinfo {volume} {75}},\ \bibinfo {pages} {35 }
  (\bibinfo {year} {1969})}\BibitemShut {NoStop}%
\bibitem [{Mon({\natexlab{a}})}]{Monitor-3}%
  \BibitemOpen
  \href@noop {} {}\bibinfo {howpublished} {\url{https://bit.ly/2ybDeMR}}
  ({\natexlab{a}}),\ \bibinfo {note} {accessed: 09/20/2018}\BibitemShut
  {NoStop}%
\bibitem [{Mon({\natexlab{b}})}]{Monitor-4}%
  \BibitemOpen
  \href@noop {} {}\bibinfo {howpublished}
  {\url{https://scintacor.com/products/6-lithium-glass}} ({\natexlab{b}}),\
  \bibinfo {note} {accessed: 09/20/2018}\BibitemShut {NoStop}%
\bibitem [{Note1()}]{Note1}%
  \BibitemOpen
  \bibinfo {note} {In this case a total of fifteen GS20\protect
  \textsuperscript {\relax \protect \fontsize {5}{6}\protect \selectfont
  \textregistered } cubic beads, 250\protect \,$\mu m$ in size, are arranged in
  a net pattern to sample the neutron beam. The light from the GS-beads is
  collected using a photo-multiplier tube and the signal processed using
  electronics developed in-house to discriminate between neutron and gamma
  events based on the signal pulse height.}\BibitemShut {Stop}%
\bibitem [{\citenamefont {Boffy}\ \emph {et~al.}(2016)\citenamefont {Boffy},
  \citenamefont {Peuget}, \citenamefont {Schweins}, \citenamefont {Beaucour},\
  and\ \citenamefont {Bermejo}}]{BOFFY201614}%
  \BibitemOpen
  \bibfield  {author} {\bibinfo {author} {\bibfnamefont {R.}~\bibnamefont
  {Boffy}}, \bibinfo {author} {\bibfnamefont {S.}~\bibnamefont {Peuget}},
  \bibinfo {author} {\bibfnamefont {R.}~\bibnamefont {Schweins}}, \bibinfo
  {author} {\bibfnamefont {J.}~\bibnamefont {Beaucour}}, \ and\ \bibinfo
  {author} {\bibfnamefont {F.}~\bibnamefont {Bermejo}},\ }\href@noop {}
  {\bibfield  {journal} {\bibinfo  {journal} {Nucl. Inst. Meth. B}\ }\textbf
  {\bibinfo {volume} {374}},\ \bibinfo {pages} {14 } (\bibinfo {year}
  {2016})}\BibitemShut {NoStop}%
\bibitem [{\citenamefont {Willendrup}\ \emph {et~al.}(2004)\citenamefont
  {Willendrup}, \citenamefont {Farhi},\ and\ \citenamefont
  {Lefmann}}]{Mcstas-paper}%
  \BibitemOpen
  \bibfield  {author} {\bibinfo {author} {\bibfnamefont {P.}~\bibnamefont
  {Willendrup}}, \bibinfo {author} {\bibfnamefont {E.}~\bibnamefont {Farhi}}, \
  and\ \bibinfo {author} {\bibfnamefont {K.}~\bibnamefont {Lefmann}},\
  }\href@noop {} {\bibfield  {journal} {\bibinfo  {journal} {Physica B}\
  }\textbf {\bibinfo {volume} {350}},\ \bibinfo {pages} {735} (\bibinfo {year}
  {2004})}\BibitemShut {NoStop}%
\bibitem [{Tob()}]{Tobyfit-ref}%
  \BibitemOpen
  \href@noop {} {\enquote {\bibinfo {title} {Tobyfit: fitting of resolution
  broadened models to data from time-of-flight chopper spectrometers},}\
  }\bibinfo {howpublished} {\url{http://tobyfit.isis.rl.ac.uk/Main_Page}},\
  \bibinfo {note} {accessed: 06/07/2018}\BibitemShut {NoStop}%
\bibitem [{\citenamefont {Russina}\ \emph {et~al.}(2012)\citenamefont
  {Russina}, \citenamefont {Mezei},\ and\ \citenamefont {Kali}}]{Russina-RRM}%
  \BibitemOpen
  \bibfield  {author} {\bibinfo {author} {\bibfnamefont {M.}~\bibnamefont
  {Russina}}, \bibinfo {author} {\bibfnamefont {F.}~\bibnamefont {Mezei}}, \
  and\ \bibinfo {author} {\bibfnamefont {G.}~\bibnamefont {Kali}},\ }\href@noop
  {} {\bibfield  {journal} {\bibinfo  {journal} {J. Phys. Conf. Ser.}\ }\textbf
  {\bibinfo {volume} {340}},\ \bibinfo {pages} {012018} (\bibinfo {year}
  {2012})}\BibitemShut {NoStop}%
\bibitem [{\citenamefont {Nakamura}\ \emph {et~al.}(2009)\citenamefont
  {Nakamura}, \citenamefont {Kajimoto}, \citenamefont {Inamura}, \citenamefont
  {Mizuno}, \citenamefont {Fujita}, \citenamefont {Yokoo},\ and\ \citenamefont
  {Arai}}]{CuGeO3_multirep}%
  \BibitemOpen
  \bibfield  {author} {\bibinfo {author} {\bibfnamefont {M.}~\bibnamefont
  {Nakamura}}, \bibinfo {author} {\bibfnamefont {R.}~\bibnamefont {Kajimoto}},
  \bibinfo {author} {\bibfnamefont {Y.}~\bibnamefont {Inamura}}, \bibinfo
  {author} {\bibfnamefont {F.}~\bibnamefont {Mizuno}}, \bibinfo {author}
  {\bibfnamefont {M.}~\bibnamefont {Fujita}}, \bibinfo {author} {\bibfnamefont
  {T.}~\bibnamefont {Yokoo}}, \ and\ \bibinfo {author} {\bibfnamefont
  {M.}~\bibnamefont {Arai}},\ }\href@noop {} {\bibfield  {journal} {\bibinfo
  {journal} {J. Phys. Soc. Japan}\ }\textbf {\bibinfo {volume} {78}},\ \bibinfo
  {pages} {093002} (\bibinfo {year} {2009})}\BibitemShut {NoStop}%
\bibitem [{Pyc()}]{Pychop-ref}%
  \BibitemOpen
  \href@noop {} {}\bibinfo {howpublished} {\url{https://bit.ly/2xXufiN}},\
  \bibinfo {note} {accessed: 08/21/2018}\BibitemShut {NoStop}%
\bibitem [{\citenamefont {Arnold}\ \emph {et~al.}(2014)\citenamefont {Arnold},
  \citenamefont {Bilheux}, \citenamefont {Borreguero}, \citenamefont {Buts},
  \citenamefont {Campbell}, \citenamefont {Chapon}, \citenamefont {Doucet},
  \citenamefont {Draper}, \citenamefont {Leal}, \citenamefont {Gigg},
  \citenamefont {Lynch}, \citenamefont {Markvardsen}, \citenamefont
  {Mikkelson}, \citenamefont {Mikkelson}, \citenamefont {Miller}, \citenamefont
  {Palmen}, \citenamefont {Parker}, \citenamefont {Passos}, \citenamefont
  {Perring}, \citenamefont {Peterson}, \citenamefont {Ren}, \citenamefont
  {Reuter}, \citenamefont {Savici}, \citenamefont {Taylor}, \citenamefont
  {Taylor}, \citenamefont {Tolchenov}, \citenamefont {Zhou},\ and\
  \citenamefont {Zikovsky}}]{Arnold2014_Mantid}%
  \BibitemOpen
  \bibfield  {author} {\bibinfo {author} {\bibfnamefont {O.}~\bibnamefont
  {Arnold}}, \bibinfo {author} {\bibfnamefont {J.}~\bibnamefont {Bilheux}},
  \bibinfo {author} {\bibfnamefont {J.}~\bibnamefont {Borreguero}}, \bibinfo
  {author} {\bibfnamefont {A.}~\bibnamefont {Buts}}, \bibinfo {author}
  {\bibfnamefont {S.}~\bibnamefont {Campbell}}, \bibinfo {author}
  {\bibfnamefont {L.}~\bibnamefont {Chapon}}, \bibinfo {author} {\bibfnamefont
  {M.}~\bibnamefont {Doucet}}, \bibinfo {author} {\bibfnamefont
  {N.}~\bibnamefont {Draper}}, \bibinfo {author} {\bibfnamefont {R.~F.}\
  \bibnamefont {Leal}}, \bibinfo {author} {\bibfnamefont {M.}~\bibnamefont
  {Gigg}}, \bibinfo {author} {\bibfnamefont {V.}~\bibnamefont {Lynch}},
  \bibinfo {author} {\bibfnamefont {A.}~\bibnamefont {Markvardsen}}, \bibinfo
  {author} {\bibfnamefont {D.}~\bibnamefont {Mikkelson}}, \bibinfo {author}
  {\bibfnamefont {R.}~\bibnamefont {Mikkelson}}, \bibinfo {author}
  {\bibfnamefont {R.}~\bibnamefont {Miller}}, \bibinfo {author} {\bibfnamefont
  {K.}~\bibnamefont {Palmen}}, \bibinfo {author} {\bibfnamefont
  {P.}~\bibnamefont {Parker}}, \bibinfo {author} {\bibfnamefont
  {G.}~\bibnamefont {Passos}}, \bibinfo {author} {\bibfnamefont
  {T.}~\bibnamefont {Perring}}, \bibinfo {author} {\bibfnamefont
  {P.}~\bibnamefont {Peterson}}, \bibinfo {author} {\bibfnamefont
  {S.}~\bibnamefont {Ren}}, \bibinfo {author} {\bibfnamefont {M.}~\bibnamefont
  {Reuter}}, \bibinfo {author} {\bibfnamefont {A.}~\bibnamefont {Savici}},
  \bibinfo {author} {\bibfnamefont {J.}~\bibnamefont {Taylor}}, \bibinfo
  {author} {\bibfnamefont {R.}~\bibnamefont {Taylor}}, \bibinfo {author}
  {\bibfnamefont {R.}~\bibnamefont {Tolchenov}}, \bibinfo {author}
  {\bibfnamefont {W.}~\bibnamefont {Zhou}}, \ and\ \bibinfo {author}
  {\bibfnamefont {J.}~\bibnamefont {Zikovsky}},\ }\href@noop {} {\bibfield
  {journal} {\bibinfo  {journal} {Nuc. Inst. Meth. A}\ }\textbf {\bibinfo
  {volume} {764}},\ \bibinfo {pages} {156 } (\bibinfo {year}
  {2014})}\BibitemShut {NoStop}%
\bibitem [{Man()}]{Mantid-website}%
  \BibitemOpen
  \href@noop {} {}\bibinfo {howpublished}
  {\url{http://www.mantidproject.org/Main_Page}},\ \bibinfo {note} {accessed:
  01/28/2019}\BibitemShut {NoStop}%
\bibitem [{\citenamefont {Ewings}\ \emph {et~al.}(2017)\citenamefont {Ewings},
  \citenamefont {Perring}, \citenamefont {Riehl-Shaw}, \citenamefont {Johnson},
  \citenamefont {Wakefield}, \citenamefont {\v{S}koro}, \citenamefont
  {Raspino}, \citenamefont {Moorby}, \citenamefont {Phillips}, \citenamefont
  {Abbley}, \citenamefont {Haynes}, \citenamefont {Waller}, \citenamefont
  {Bewley},\ and\ \citenamefont {Stewart}}]{MAPS-FluxLoss}%
  \BibitemOpen
  \bibfield  {author} {\bibinfo {author} {\bibfnamefont {R.~A.}\ \bibnamefont
  {Ewings}}, \bibinfo {author} {\bibfnamefont {T.~G.}\ \bibnamefont {Perring}},
  \bibinfo {author} {\bibfnamefont {R.~C.}\ \bibnamefont {Riehl-Shaw}},
  \bibinfo {author} {\bibfnamefont {E.~L.}\ \bibnamefont {Johnson}}, \bibinfo
  {author} {\bibfnamefont {S.~R.}\ \bibnamefont {Wakefield}}, \bibinfo {author}
  {\bibfnamefont {G.}~\bibnamefont {\v{S}koro}}, \bibinfo {author}
  {\bibfnamefont {D.}~\bibnamefont {Raspino}}, \bibinfo {author} {\bibfnamefont
  {S.~R.}\ \bibnamefont {Moorby}}, \bibinfo {author} {\bibfnamefont
  {P.}~\bibnamefont {Phillips}}, \bibinfo {author} {\bibfnamefont {D.~D.}\
  \bibnamefont {Abbley}}, \bibinfo {author} {\bibfnamefont {D.~J.}\
  \bibnamefont {Haynes}}, \bibinfo {author} {\bibfnamefont {S.~P.}\
  \bibnamefont {Waller}}, \bibinfo {author} {\bibfnamefont {R.~I.}\
  \bibnamefont {Bewley}}, \ and\ \bibinfo {author} {\bibfnamefont {J.~R.}\
  \bibnamefont {Stewart}},\ }\href@noop {} {\bibfield  {journal} {\bibinfo
  {journal} {RAL Technical Report}\ }\textbf {\bibinfo {volume} {007}}
  (\bibinfo {year} {2017})}\BibitemShut {NoStop}%
\bibitem [{nge()}]{ngem-ref}%
  \BibitemOpen
  \href@noop {} {}\bibinfo {howpublished}
  {\url{https://www.bbtech.co.jp/en/}},\ \bibinfo {note} {accessed:
  06/07/2018}\BibitemShut {NoStop}%
\bibitem [{\citenamefont {Lipscombe}\ \emph {et~al.}(2007)\citenamefont
  {Lipscombe}, \citenamefont {Hayden}, \citenamefont {Vignolle}, \citenamefont
  {McMorrow},\ and\ \citenamefont {Perring}}]{Lipscombe-PRL07}%
  \BibitemOpen
  \bibfield  {author} {\bibinfo {author} {\bibfnamefont {O.~J.}\ \bibnamefont
  {Lipscombe}}, \bibinfo {author} {\bibfnamefont {S.~M.}\ \bibnamefont
  {Hayden}}, \bibinfo {author} {\bibfnamefont {B.}~\bibnamefont {Vignolle}},
  \bibinfo {author} {\bibfnamefont {D.~F.}\ \bibnamefont {McMorrow}}, \ and\
  \bibinfo {author} {\bibfnamefont {T.~G.}\ \bibnamefont {Perring}},\
  }\href@noop {} {\bibfield  {journal} {\bibinfo  {journal} {Phys. Rev. Lett.}\
  }\textbf {\bibinfo {volume} {99}},\ \bibinfo {pages} {067002} (\bibinfo
  {year} {2007})}\BibitemShut {NoStop}%
\bibitem [{Note2()}]{Note2}%
  \BibitemOpen
  \bibinfo {note} {$\psi = 0$ refers to the case when the incident beam is
  parallel to the $(1,1,0)$-direction}\BibitemShut {NoStop}%
\bibitem [{\citenamefont {Fujita}\ \emph {et~al.}(2013)\citenamefont {Fujita},
  \citenamefont {Frost}, \citenamefont {Bennington}, \citenamefont {Kajimoto},
  \citenamefont {Nakamura}, \citenamefont {Inamura}, \citenamefont {Mizuno},
  \citenamefont {Ikeuchi},\ and\ \citenamefont {Arai}}]{CuGeO3-4seasons}%
  \BibitemOpen
  \bibfield  {author} {\bibinfo {author} {\bibfnamefont {M.}~\bibnamefont
  {Fujita}}, \bibinfo {author} {\bibfnamefont {C.~D.}\ \bibnamefont {Frost}},
  \bibinfo {author} {\bibfnamefont {S.~M.}\ \bibnamefont {Bennington}},
  \bibinfo {author} {\bibfnamefont {R.}~\bibnamefont {Kajimoto}}, \bibinfo
  {author} {\bibfnamefont {M.}~\bibnamefont {Nakamura}}, \bibinfo {author}
  {\bibfnamefont {Y.}~\bibnamefont {Inamura}}, \bibinfo {author} {\bibfnamefont
  {F.}~\bibnamefont {Mizuno}}, \bibinfo {author} {\bibfnamefont
  {K.}~\bibnamefont {Ikeuchi}}, \ and\ \bibinfo {author} {\bibfnamefont
  {M.}~\bibnamefont {Arai}},\ }\href@noop {} {\bibfield  {journal} {\bibinfo
  {journal} {J. Phys. Soc. Jpn.}\ }\textbf {\bibinfo {volume} {82}},\ \bibinfo
  {pages} {084708} (\bibinfo {year} {2013})}\BibitemShut {NoStop}%
\bibitem [{\citenamefont {Arai}\ \emph {et~al.}(2013)\citenamefont {Arai},
  \citenamefont {Kajimoto}, \citenamefont {Nakamura}, \citenamefont {Inamura},
  \citenamefont {Nakajima}, \citenamefont {Shibata}, \citenamefont {Takahashi},
  \citenamefont {Suzuki}, \citenamefont {Takata}, \citenamefont {Yamada},\ and\
  \citenamefont {Itoh}}]{Arai-CuGeO3-overview}%
  \BibitemOpen
  \bibfield  {author} {\bibinfo {author} {\bibfnamefont {M.}~\bibnamefont
  {Arai}}, \bibinfo {author} {\bibfnamefont {R.}~\bibnamefont {Kajimoto}},
  \bibinfo {author} {\bibfnamefont {M.}~\bibnamefont {Nakamura}}, \bibinfo
  {author} {\bibfnamefont {Y.}~\bibnamefont {Inamura}}, \bibinfo {author}
  {\bibfnamefont {K.}~\bibnamefont {Nakajima}}, \bibinfo {author}
  {\bibfnamefont {K.}~\bibnamefont {Shibata}}, \bibinfo {author} {\bibfnamefont
  {N.}~\bibnamefont {Takahashi}}, \bibinfo {author} {\bibfnamefont
  {J.}~\bibnamefont {Suzuki}}, \bibinfo {author} {\bibfnamefont
  {S.}~\bibnamefont {Takata}}, \bibinfo {author} {\bibfnamefont
  {T.}~\bibnamefont {Yamada}}, \ and\ \bibinfo {author} {\bibfnamefont
  {S.}~\bibnamefont {Itoh}},\ }\href@noop {} {\bibfield  {journal} {\bibinfo
  {journal} {J. Phys. Soc. Japan}\ }\textbf {\bibinfo {volume} {82}},\ \bibinfo
  {pages} {SA024} (\bibinfo {year} {2013})}\BibitemShut {NoStop}%
\bibitem [{\citenamefont {Kajimoto}\ \emph
  {et~al.}(2013{\natexlab{a}})\citenamefont {Kajimoto}, \citenamefont
  {Nakamura}, \citenamefont {Inamura}, \citenamefont {Ikeuchi}, \citenamefont
  {Ji}, \citenamefont {Nakajima}, \citenamefont {Ohira-Kawamura}, \citenamefont
  {Kambara}, \citenamefont {Sawabe}, \citenamefont {Kamiya}, \citenamefont
  {Inoue}, \citenamefont {Futagami}, \citenamefont {Kobayashi}, \citenamefont
  {Kishi}, \citenamefont {Satou}, \citenamefont {Aizawa},\ and\ \citenamefont
  {Arai}}]{Kajimoto-CuGeO3}%
  \BibitemOpen
  \bibfield  {author} {\bibinfo {author} {\bibfnamefont {R.}~\bibnamefont
  {Kajimoto}}, \bibinfo {author} {\bibfnamefont {M.}~\bibnamefont {Nakamura}},
  \bibinfo {author} {\bibfnamefont {Y.}~\bibnamefont {Inamura}}, \bibinfo
  {author} {\bibfnamefont {K.}~\bibnamefont {Ikeuchi}}, \bibinfo {author}
  {\bibfnamefont {S.}~\bibnamefont {Ji}}, \bibinfo {author} {\bibfnamefont
  {K.}~\bibnamefont {Nakajima}}, \bibinfo {author} {\bibfnamefont
  {S.}~\bibnamefont {Ohira-Kawamura}}, \bibinfo {author} {\bibfnamefont
  {W.}~\bibnamefont {Kambara}}, \bibinfo {author} {\bibfnamefont
  {M.}~\bibnamefont {Sawabe}}, \bibinfo {author} {\bibfnamefont
  {A.}~\bibnamefont {Kamiya}}, \bibinfo {author} {\bibfnamefont
  {K.}~\bibnamefont {Inoue}}, \bibinfo {author} {\bibfnamefont
  {T.}~\bibnamefont {Futagami}}, \bibinfo {author} {\bibfnamefont
  {M.}~\bibnamefont {Kobayashi}}, \bibinfo {author} {\bibfnamefont
  {A.}~\bibnamefont {Kishi}}, \bibinfo {author} {\bibfnamefont
  {K.}~\bibnamefont {Satou}}, \bibinfo {author} {\bibfnamefont
  {K.}~\bibnamefont {Aizawa}}, \ and\ \bibinfo {author} {\bibfnamefont
  {M.}~\bibnamefont {Arai}},\ }\href@noop {} {\bibfield  {journal} {\bibinfo
  {journal} {J. Phys. Soc. Japan}\ }\textbf {\bibinfo {volume} {82}},\ \bibinfo
  {pages} {SA032} (\bibinfo {year} {2013}{\natexlab{a}})}\BibitemShut {NoStop}%
\bibitem [{\citenamefont {Kajimoto}\ \emph
  {et~al.}(2013{\natexlab{b}})\citenamefont {Kajimoto}, \citenamefont
  {Nakamura}, \citenamefont {Nakajima},\ and\ \citenamefont
  {Fujita}}]{Kajimoto-NIMA}%
  \BibitemOpen
  \bibfield  {author} {\bibinfo {author} {\bibfnamefont {R.}~\bibnamefont
  {Kajimoto}}, \bibinfo {author} {\bibfnamefont {M.}~\bibnamefont {Nakamura}},
  \bibinfo {author} {\bibfnamefont {K.}~\bibnamefont {Nakajima}}, \ and\
  \bibinfo {author} {\bibfnamefont {M.}~\bibnamefont {Fujita}},\ }\href@noop {}
  {\bibfield  {journal} {\bibinfo  {journal} {Nucl. Inst. Meth. A}\ }\textbf
  {\bibinfo {volume} {729}},\ \bibinfo {pages} {365 } (\bibinfo {year}
  {2013}{\natexlab{b}})}\BibitemShut {NoStop}%
\bibitem [{\citenamefont {Ewings}\ \emph {et~al.}(2016)\citenamefont {Ewings},
  \citenamefont {Buts}, \citenamefont {Le}, \citenamefont {van Duijn},
  \citenamefont {Bustinduy},\ and\ \citenamefont {Perring}}]{Ewings-Horace}%
  \BibitemOpen
  \bibfield  {author} {\bibinfo {author} {\bibfnamefont {R.~A.}\ \bibnamefont
  {Ewings}}, \bibinfo {author} {\bibfnamefont {A.}~\bibnamefont {Buts}},
  \bibinfo {author} {\bibfnamefont {M.~D.}\ \bibnamefont {Le}}, \bibinfo
  {author} {\bibfnamefont {J.}~\bibnamefont {van Duijn}}, \bibinfo {author}
  {\bibfnamefont {I.}~\bibnamefont {Bustinduy}}, \ and\ \bibinfo {author}
  {\bibfnamefont {T.~G.}\ \bibnamefont {Perring}},\ }\href@noop {} {\bibfield
  {journal} {\bibinfo  {journal} {Nucl. Inst. Meth. A}\ }\textbf {\bibinfo
  {volume} {834}},\ \bibinfo {pages} {132 } (\bibinfo {year}
  {2016})}\BibitemShut {NoStop}%
\bibitem [{\citenamefont {Caux}\ and\ \citenamefont
  {Hagemans}()}]{Caux-JStatMech}%
  \BibitemOpen
  \bibfield  {author} {\bibinfo {author} {\bibfnamefont {J.-S.}\ \bibnamefont
  {Caux}}\ and\ \bibinfo {author} {\bibfnamefont {R.}~\bibnamefont
  {Hagemans}},\ }\href@noop {} {\bibfield  {journal} {\bibinfo  {journal} {J.
  Stat. Mech.}\ }\textbf {\bibinfo {volume} {2006}},\ \bibinfo {pages}
  {P12013}}\BibitemShut {NoStop}%
\bibitem [{\citenamefont {Babkevich}\ \emph {et~al.}(2017)\citenamefont
  {Babkevich}, \citenamefont {Shaik}, \citenamefont
  {Lan\ifmmode~\mbox{\c{c}}\else \c{c}\fi{}on}, \citenamefont {Kikkawa},
  \citenamefont {Enderle}, \citenamefont {Ewings}, \citenamefont {Walker},
  \citenamefont {Adroja}, \citenamefont {Manuel}, \citenamefont {Khalyavin},
  \citenamefont {Taguchi}, \citenamefont {Tokura}, \citenamefont {Soda},
  \citenamefont {Masuda},\ and\ \citenamefont
  {R\o{}nnow}}]{Babbers-Cu-cupolas}%
  \BibitemOpen
  \bibfield  {author} {\bibinfo {author} {\bibfnamefont {P.}~\bibnamefont
  {Babkevich}}, \bibinfo {author} {\bibfnamefont {N.~E.}\ \bibnamefont
  {Shaik}}, \bibinfo {author} {\bibfnamefont {D.}~\bibnamefont
  {Lan\ifmmode~\mbox{\c{c}}\else \c{c}\fi{}on}}, \bibinfo {author}
  {\bibfnamefont {A.}~\bibnamefont {Kikkawa}}, \bibinfo {author} {\bibfnamefont
  {M.}~\bibnamefont {Enderle}}, \bibinfo {author} {\bibfnamefont {R.~A.}\
  \bibnamefont {Ewings}}, \bibinfo {author} {\bibfnamefont {H.~C.}\
  \bibnamefont {Walker}}, \bibinfo {author} {\bibfnamefont {D.~T.}\
  \bibnamefont {Adroja}}, \bibinfo {author} {\bibfnamefont {P.}~\bibnamefont
  {Manuel}}, \bibinfo {author} {\bibfnamefont {D.~D.}\ \bibnamefont
  {Khalyavin}}, \bibinfo {author} {\bibfnamefont {Y.}~\bibnamefont {Taguchi}},
  \bibinfo {author} {\bibfnamefont {Y.}~\bibnamefont {Tokura}}, \bibinfo
  {author} {\bibfnamefont {M.}~\bibnamefont {Soda}}, \bibinfo {author}
  {\bibfnamefont {T.}~\bibnamefont {Masuda}}, \ and\ \bibinfo {author}
  {\bibfnamefont {H.~M.}\ \bibnamefont {R\o{}nnow}},\ }\href@noop {} {\bibfield
   {journal} {\bibinfo  {journal} {Phys. Rev. B}\ }\textbf {\bibinfo {volume}
  {96}},\ \bibinfo {pages} {014410} (\bibinfo {year} {2017})}\BibitemShut
  {NoStop}%
\bibitem [{\citenamefont {Ewings}\ \emph {et~al.}(2011)\citenamefont {Ewings},
  \citenamefont {Perring}, \citenamefont {Gillett}, \citenamefont {Das},
  \citenamefont {Sebastian}, \citenamefont {Taylor}, \citenamefont {Guidi},\
  and\ \citenamefont {Boothroyd}}]{Ewings-Sr122}%
  \BibitemOpen
  \bibfield  {author} {\bibinfo {author} {\bibfnamefont {R.~A.}\ \bibnamefont
  {Ewings}}, \bibinfo {author} {\bibfnamefont {T.~G.}\ \bibnamefont {Perring}},
  \bibinfo {author} {\bibfnamefont {J.}~\bibnamefont {Gillett}}, \bibinfo
  {author} {\bibfnamefont {S.~D.}\ \bibnamefont {Das}}, \bibinfo {author}
  {\bibfnamefont {S.~E.}\ \bibnamefont {Sebastian}}, \bibinfo {author}
  {\bibfnamefont {A.~E.}\ \bibnamefont {Taylor}}, \bibinfo {author}
  {\bibfnamefont {T.}~\bibnamefont {Guidi}}, \ and\ \bibinfo {author}
  {\bibfnamefont {A.~T.}\ \bibnamefont {Boothroyd}},\ }\href@noop {} {\bibfield
   {journal} {\bibinfo  {journal} {Phys. Rev. B}\ }\textbf {\bibinfo {volume}
  {83}},\ \bibinfo {pages} {214519} (\bibinfo {year} {2011})}\BibitemShut
  {NoStop}%
\bibitem [{\citenamefont {Pinna}\ \emph {et~al.}(2018)\citenamefont {Pinna},
  \citenamefont {Rudic}, \citenamefont {Parker}, \citenamefont {Armstrong},
  \citenamefont {Zanetti}, \citenamefont {Škoro}, \citenamefont {Waller},
  \citenamefont {Zacek}, \citenamefont {Smith}, \citenamefont {Capstick},
  \citenamefont {McPhail}, \citenamefont {Pooley}, \citenamefont {Howells},
  \citenamefont {Gorini},\ and\ \citenamefont
  {Fernandez-Alonso}}]{Tosca-paper}%
  \BibitemOpen
  \bibfield  {author} {\bibinfo {author} {\bibfnamefont {R.~S.}\ \bibnamefont
  {Pinna}}, \bibinfo {author} {\bibfnamefont {S.}~\bibnamefont {Rudic}},
  \bibinfo {author} {\bibfnamefont {S.~F.}\ \bibnamefont {Parker}}, \bibinfo
  {author} {\bibfnamefont {J.}~\bibnamefont {Armstrong}}, \bibinfo {author}
  {\bibfnamefont {M.}~\bibnamefont {Zanetti}}, \bibinfo {author} {\bibfnamefont
  {G.}~\bibnamefont {Škoro}}, \bibinfo {author} {\bibfnamefont {S.~P.}\
  \bibnamefont {Waller}}, \bibinfo {author} {\bibfnamefont {D.}~\bibnamefont
  {Zacek}}, \bibinfo {author} {\bibfnamefont {C.~A.}\ \bibnamefont {Smith}},
  \bibinfo {author} {\bibfnamefont {M.~J.}\ \bibnamefont {Capstick}}, \bibinfo
  {author} {\bibfnamefont {D.~J.}\ \bibnamefont {McPhail}}, \bibinfo {author}
  {\bibfnamefont {D.~E.}\ \bibnamefont {Pooley}}, \bibinfo {author}
  {\bibfnamefont {G.~D.}\ \bibnamefont {Howells}}, \bibinfo {author}
  {\bibfnamefont {G.}~\bibnamefont {Gorini}}, \ and\ \bibinfo {author}
  {\bibfnamefont {F.}~\bibnamefont {Fernandez-Alonso}},\ }\href@noop {}
  {\bibfield  {journal} {\bibinfo  {journal} {Nuc. Inst. Meth. A}\ }\textbf
  {\bibinfo {volume} {896}},\ \bibinfo {pages} {68 } (\bibinfo {year}
  {2018})}\BibitemShut {NoStop}%
\bibitem [{\citenamefont {Shirane}\ \emph {et~al.}(2006)\citenamefont
  {Shirane}, \citenamefont {Shapiro},\ and\ \citenamefont
  {Tranquada}}]{Tranquada-book}%
  \BibitemOpen
  \bibfield  {author} {\bibinfo {author} {\bibfnamefont {G.}~\bibnamefont
  {Shirane}}, \bibinfo {author} {\bibfnamefont {S.~M.}\ \bibnamefont
  {Shapiro}}, \ and\ \bibinfo {author} {\bibfnamefont {J.~M.}\ \bibnamefont
  {Tranquada}},\ }\href@noop {} {\emph {\bibinfo {title} {Neutron Scattering
  with a Triple-Axis spectrometer}}}\ (\bibinfo  {publisher} {Cambridge},\
  \bibinfo {year} {2006})\BibitemShut {NoStop}%
\end{thebibliography}%

\end{document}